%% file: main.tex
\documentclass[screen,nonacm]{acmart}

\usepackage{framed}
\usepackage{tabularx}
\usepackage{subfig}
\usepackage{tikz} 

\newcommand\todo[1]{{\color{black}{#1}}}%
\newcommand\ok[1]{{\color{black}{#1}}}%
\newcommand\rev[1]{{\color{black}{#1}}}%
\setcopyright{none} 
\renewcommand\footnotetextcopyrightpermission[1]{}
\setcopyright{rightsretained}
\copyrightyear{2026}
\acmYear{2026}
\acmDOI{XXXXXXX.XXXXXXX}
\begin{document}	
\title[Prompt Engineer]{Prompt Engineer: Analyzing Hard and Soft Skill Requirements in the AI Job Market}

\author{An Vu}
\email{an.vu@student.oulu.fi}
\affiliation{%
	\institution{University of Oulu}
	\city{Oulu}
	\country{Finland}
}

\author{Jonas Oppenlaender}
\email{jonas.oppenlaender@oulu.fi}
\affiliation{%
	\institution{University of Oulu}
	\city{Oulu}
	\country{Finland}
}


\begin{abstract}
%
%
The rise of large language models (LLMs) has prompted speculation about a new job role: the prompt engineer.
However, empirical evidence on how common such positions are and what skills employers expect of them remains scarce.
In this paper, we present a data-driven analysis of a snapshot of LinkedIn job postings for data and AI roles taken in April 2025, after LLMs had entered everyday
work but before LLM-based agents became widely adopted.
We analyzed 20,662~job postings on LinkedIn for prompt engineers (72 postings) and four established data and AI roles: data scientist, data engineer, data analyst, and machine learning engineer.
We find that prompt engineer roles were rare even at this point in time, making up less than 0.5\% of the postings across the five roles we sampled.
Because our snapshot predates the widespread adoption of LLM-based agents, we argue that the data captures the role of prompt engineer under favorable conditions. Whether its share has since declined is an open question that later snapshots can test.
Despite its rarity, prompt engineering has a distinct skill profile. Prompt engineer postings emphasize AI knowledge (22.8\%), communication
(21.9\%), prompt design (18.7\%), and creative problem-solving (15.8\%).
These soft and hard skill requirements differ significantly from those of the four established roles.
Taken together, the distinct skill profile and the scarcity of dedicated postings suggest that prompt engineering is best understood not as a profession of its own, but as a distinct skill that is useful across many professions.
Our findings inform job seekers, employers, and educators about how to position prompt engineering in the evolving AI workforce.
\end{abstract}




\maketitle


\section{Introduction}
\label{sec:introduction}
The rise of large language models (LLMs) has been transforming various industries \citep{10822885}.
AI enables advancements in automation, content generation, and AI-driven decision-making.
However, the effectiveness of AI models heavily depends on how prompts are designed and structured \citep{Oppenlaender_2023}, giving rise to the practice of \textit{prompt engineering}.
Prompts are programs that steer the LLM without changing its weights \citep{karpathy}.
Not long after the introduction of ChatGPT, some employers began recruiting \textit{prompt engineers} to get the most out of LLMs.
For instance, in July 2024, the AI Safety Institute advertised a ``Senior Prompt Engineer'' in its evaluations team, with a salary of 
\pounds 65,000--\pounds 135,000. The role involved eliciting capabilities from advanced AI systems and improving agent scaffolds, and it 
required experience with prompting techniques such as chain-of-thought and ReAct \citep{webarchive}.
Yet, it remains unclear how common such positions are and what they entail. There is no standardized definition of a prompt engineer, leaving the scope, required skills, and industry relevance of this emerging job role uncertain.
%
Due to the timely relevance of LLM applications in industry, it is important to develop an understanding of the required knowledge and skills for this emerging job role.
However, because of the novelty of AI technology, individuals who are interested in pursuing the role of the prompt engineer may face challenges in understanding the qualifications required.
Therefore, addressing this gap is beneficial to both job seekers and practitioners in industry as well as employers who want to identify suitable candidates.

In this paper, we systematically analyze job advertisements published on LinkedIn using a text mining approach to identify the knowledge areas and skills that employers expect of prompt engineers.
Our data is a snapshot of LinkedIn job postings collected on April 7, 2025. This date is significant. By then, LLMs had become part of everyday work, but LLM-based agents entered mainstream use only in late 2025.
People still wrote prompts largely by hand, so if prompt engineering ever had a strong case as a dedicated job role, it was at this time.
Therefore, the snapshot lets us test whether employers valued prompt engineering enough to create dedicated positions for it.

To position prompt engineering in the job market, we compare it against four established data and AI roles: data analyst, data scientist, data engineer, and machine learning engineer.
We 
exclude the \rev{title of AI engineer. Although this title predates the prompt engineer, employers use it for heterogeneous profiles \citep{10.1145/3522664.3528594}, and since the rise of LLMs its scope has arguably come to overlap with prompt engineering, so it could not serve as a stable reference point.}
The four roles serve as a reference frame. If prompt engineering did not become a profession of its own, its responsibilities would most plausibly be folded into these roles. If, in contrast, prompt engineering stood as a distinct job role on its own, we would expect it to have a skill profile that is distinct from these roles.
%

Using this snapshot, we seek to answer the following research questions: 
\begin{itemize}%
	\item
	RQ1: \textit{How prevalent are prompt engineer positions compared to more established job roles?} 
	\item
	RQ2: \textit{What soft skills are most valued for prompt engineers, and how do these differ from other job roles?} 
	\item
	RQ3: \textit{What technical skills define the prompt engineer role, and how do these hard skills create a distinct professional profile?}
\end{itemize}

By analyzing job postings from this point in time, this study documents how employers defined the prompt engineer role under conditions that favored dedicated prompt engineering positions, and what this implies for job seekers and employers as LLM-based agents take over more of the prompting work.%

The remainder of this paper is structured as follows.
We review related literature on prompt engineering and the extraction of skills in Section \ref{sec:relatedwork}.
We then describe our approach to collecting and extracting information from job descriptions in Section~\ref{sec:method}.
We present our findings in Section~\ref{sec:results}, contrasting the skills of prompt engineers with those of the four reference roles.
We discuss our findings in Section \ref{sec:discussion} and conclude in Section \ref{sec:conclusion}.

\section{Related Work}%
\label{sec:relatedwork}%
\subsection{The Emergence of the Prompt Engineer}%
Artificial intelligence (AI) is changing the way workers perform their jobs and the skills they require \citep{OECD,Brasse2024}.
Generative artificial intelligence has a strong impact on the future of work \citep{Cazzaniga2024Gen}. 
While concerns and anxiety over automation are growing \citep{Dinnen2022}, 
AI also creates new professional opportunities \citep{kutela2023ai}.
There is a demand for a new digital workforce with a novel set of AI literacy \citep{3313831.3376727.pdf} and digital skills \citep{VANLAAR2017577}.
One such novel job is the prompt engineer.

The emergence of LLMs and `prompt engineering' (i.e., the practice of 
optimizing input prompts for LLMs) has sparked interest among researchers investigating how prompt engineering is applied in the job market, and the qualifications required for this emerging job role.

\citet{kutela_artificial_2023}
examined the nature, qualifications, and compensation of prompt engineers through an analysis of LinkedIn job descriptions. Their study mainly focused on collocated keywords and frequency counts using statistical approaches. While these methods are well established and documented, they may fail to capture higher-level semantic meaning in the text and struggle to answer more complex questions about the profession's requirements and expectations.

%
\citet{george_exploring_2023} used a qualitative method of interviewing industry experts, university researchers, and students to examine opportunities for prompt engineers, including job requirements and the characteristics of these emerging roles.
Although this approach is appropriate for the research questions, they provide no transparency regarding their interview protocols, participant selection criteria, or analytical framework, making their findings hard to replicate.%

\rev{%
A closely related line of work has examined the AI engineer, a job title that predates the prompt engineer.
\citet{10.1145/3522664.3528594} analyzed 367~job ads for AI engineers in the Netherlands, published between April 2018 and April 2021, and manually coded 125 of them.
They found that the tasks of AI engineers span five categories: business understanding, data engineering, modeling, software development, and operations engineering.
Employers used the same title for three distinct profiles: a data science engineer focused on modeling, an AI software engineer focused on software development, and a generalist AI engineer covering both.
The authors concluded that the definition of an AI engineer is less established than that of a data scientist or a software engineer.
Our study applies a similar job-ad-based approach to the prompt engineer, and it builds on their finding when selecting the roles we compare prompt engineering against (see Section~\ref{sec:method}).
}%

With the rise of LLMs, significantly more researchers have approved and applied LLMs to qualitative research \citep{morgan_exploring_2023,kirsten_assistance_2025,zhang_redefining_2024}.
%
Several studies (e.g., \citet{kirsten_assistance_2025,zhang_redefining_2024}) provide their own frameworks and approaches to effectively apply LLMs to qualitative research methods. This paper takes a mix of some of the suggested approaches and adapts them to our specific needs.%


There is 
consensus among 
academic researchers,
online communities, and industry professionals
on the 
perishable nature of prompt engineering skills.
\citet{muktadir2023brief} researched the historical development of prompting, which shows the rapid evolution of prompt engineering and the need for continuous adaptation of skills in this domain.
In the following section, we review approaches for extracting skills from job postings.%
\subsection{Techniques for Skill Extraction}%
The analysis of prompt engineering skills requires robust methodologies to identify the skills from job descriptions, a task that involves information extraction (IE) and natural language processing (NLP). This analysis faces multiple challenges, stemming from the  lack of high-quality data about the nature of work \cite{6531.full.pdf}, the unstructured nature of job descriptions \citep{acta}, the inherent limitation of information retrieval and information extraction techniques \cite{doi:10.1177/1847979019890771,Manning}, and the evolving terminology within the \todo{prompt engineering field} \cite{schulhoff2025promptreportsystematicsurvey}.

\citeauthor{kivimaki2013graph}
suggest a method called Elisit that uses classical information retrieval techniques \cite{kivimaki2013graph}. The method consists of two phases. First, similarity matching is used to map input text to Wikipedia articles using TF-IDF or other similarity measures. The method then applies a modified spreading activation algorithm on Wikipedia's hyperlink graph to identify skill nodes.
This approach would work with more established fields, because it relies on Wikipedia to have  
sufficient information.
However, this approach may not be as effective for analyzing prompt engineering, an emergent field that is still largely unexplored and potentially underrepresented in established knowledge bases, such as Wikipedia.

\citeauthor{peerj-cs-121.pdf} presented ScholarLens \cite{peerj-cs-121.pdf}, a system for automatically constructing user profiles from scholarly literature, including competences.
Their approach applies techniques from natural language processing (NLP)  and linked open data entity linking.

\citeauthor{zhang-etal-2022-kompetencer}
presented Kompetencer \cite{zhang-etal-2022-kompetencer}, a Danish dataset of job postings with nested spans of competences, resulting from classifying job competences from job postings.
The authors created this dataset by using the API of the European Skills, Competences, Qualifications and Occupations (ESCO; \citet{ESCO}) taxonomy to obtain fine-grained labels via distant supervision.
However, this taxonomy does not captured the novel set of skills required in the emerging role of prompt engineering.

\citeauthor{Calanca2019}
analyzed soft skills in job advertisements by drawing on a semi-automatic approach based on crowdsourcing and text mining for extracting a list of soft skills \cite{Calanca2019}.
Our work differs in that we do not employ crowdsourcing and also extract hard skills.

\citeauthor{michalczyk_demystifying_2021}
specifically investigated the data scientist job role, by processing 25,104 online job postings with a text mining approach combining topic modeling, clustering, and expert assessment \cite{michalczyk_demystifying_2021}.
The authors identified six  job roles in data science in three major knowledge domains.


\citeauthor{zhang2022skillspanhardsoftskill} applied an approach based on sequence labeling \cite{zhang2022skillspanhardsoftskill}. Their model, SkillSpan, is a BERT-based model that is fine-tuned on manually annotated job postings.
Their approach reached 57.7\% F1 score, and their approach also includes transparent annotation guidelines \citep{zhang2022skillspanhardsoftskill} and was later evaluated by other researchers (e.g., \citet{nguyen2024rethinkingskillextractionjob}).
However, their approach faces a critical limitation: because of the way they annotated skills, the extracted skills often include long spans of text containing extraneous information. 
However, the identified skills, while being relevant, include contextual elements complicating downstream analysis such as skill categorization.%

To overcome the limitations in the above approaches, many researchers have turned to pretrained language models, as reviewed in the following section.

\subsection{Skill Extraction with LLMs}%
Recent research has shifted toward leveraging language models for skill extraction from unstructured text \citep{nguyen2024rethinkingskillextractionjob,herandi2024skillllmrepurposinggeneralpurposellms}, offering promising alternatives to traditional sequence labeling approaches. 
Notable among these approaches is the work by  \citet{herandi2024skillllmrepurposinggeneralpurposellms}, who
fine-tuned an 8B-parameter LLaMA model~\citep{touvron2023llamaopenefficientfoundation} on the SkillSpan dataset~\citep{zhang2022skillspanhardsoftskill}, achieving a 64.8\% F1 score using their specialized prompting techniques.


\citeauthor{zora230653} presented an approach for extracting  and classifying  skill requirements in German-speaking job postings \cite{zora230653}.
Their method adopts pretrained transformer-based language models for extracting skills, using context from job postings and the ESCO taxonomy \citep{ESCO}.
As with Kompetencer \citep{zhang-etal-2022-kompetencer}, the approach is limited as it draws on an existing domain taxonomy to match potentially novel skills.

Large language models have also peaked interest in research into leveraging LLMs in qualitative research.
\citeauthor{morgan_exploring_2023} has examined the effectiveness of using ChatGPT in qualitative analysis \cite{morgan_exploring_2023}. While the language model is less successful at locating subtle themes, it is more successful at reproducing concrete, descriptive themes.
\citeauthor{kirsten_assistance_2025} conducted interviews with 15~HCI researchers experienced in qualitative data analysis to understand their concerns about leveraging LLMs in research \cite{kirsten_assistance_2025}. Based on the research, they suggest frameworks to leverage the strengths of LLMs while minimizing problems that come with LLM use.
\citeauthor{zhang_redefining_2024} further advanced this field by developing a comprehensive framework for designing prompts to enhance ChatGPT's effectiveness in thematic analysis \cite{zhang_redefining_2024}. Through semi-structured interviews with seventeen participants and collaboration with thirteen qualitative researchers, they identified that improving transparency, providing guidance on prompts, and strengthening users' understanding of LLMs' capabilities significantly enhanced researchers' ability to effectively utilize ChatGPT for qualitative analysis.

This paper adopts an LLM-based approach to extract skills from unstructured text due to its demonstrated effectiveness
. By combining manual annotation with LLM-assisted qualitative analysis, as suggested by \citet{kirsten_assistance_2025}, and incorporating prompt design principles from \citet{zhang_redefining_2024}, we aim to overcome the limitations of traditional skill extraction techniques in analyzing the emerging field of prompt engineering.
In the following section, we describe our approach to extracting skills from job postings.%
\section{Method}%
\label{sec:method}%
\subsection{Data Collection}
For this study, we use a dataset of job postings collected from LinkedIn on April 7, 2025.
The one-day crawl provides us a snapshot view into the online job market on this given date.

%
%
We argue that this date is not a limitation, but a useful vantage point.
By April 2025, the initial novelty of ChatGPT had worn off, and many professionals relied heavily on conversational AI and
LLMs in their daily work.
LLM-based agents 
that reliably follow instructions
and complete multi-step tasks over long horizons became practical for everyday use only in late 2025. The release of Claude Opus 4.5 in
November 2025 is often cited as the start of this ``agentic era.''
Our snapshot is dated before this agentic era, at a time when fully agentic workflows were not even imaginable by many users.
The snapshot captures the labor market about six months before
agents entered mainstream use. At that time, people interacted with LLMs mostly by writing prompts by hand, without fully automated AI
assistants and coding agents to do it for them.
If prompt engineering ever had a strong case as a dedicated job role, it was at this moment.
This lets us use the snapshot as a test of how much value companies placed on prompt engineering.
If we find many job postings for prompt engineers, companies clearly considered the skill valuable enough to create dedicated roles for it. If we find only a few, then even under
conditions that favored prompt engineering, companies did not treat it as a standalone profession. Instead, they treated it as a skill
expected within existing roles. Because agents increasingly write and refine their own prompts, we expect demand for dedicated prompt
engineers to have declined since our data collection.
Our findings should, therefore, be read as a likely peak in demand for dedicated prompt engineering roles rather than as a baseline.

While job postings on LinkedIn do not reflect all jobs advertised online, LinkedIn represents one of the largest global platforms for professional networking and job advertisements, making it a valuable source for analyzing current labor market trends.
The data was crawled across the 60~countries with the most technological expertise, based on the U.S. News \& World Report rankings \citep{usnews_tech_expertise_2025}.
\todo{The countries are listed in the supplemental material.}
The distribution of jobs reflects the concentration of tech jobs world-wide.
\rev{%
For our comparative analysis, we selected four established data and AI roles as a reference frame: data analyst, data scientist, data engineer, and machine learning engineer.
Our aim was not to identify the closest neighbor of the prompt engineer, but to test whether prompt engineering constitutes a distinct skill profile relative to the established professions into which LLM-related work would most plausibly be folded. If prompt engineering was absorbed by existing roles, its skill profile should closely resemble one of these roles. If it stood on its own, its profile should differ from all of them.
These four roles have well-documented and comparatively stable skill profiles (e.g., \citep{michalczyk_demystifying_2021}), and together they span a spectrum from business-facing analysis (data analyst) to technical model development (machine learning engineer).
\rev{The title ``AI engineer'' was not included as a baseline, although it predates the prompt engineer.
In their analysis of 367~job ads for AI engineers, \citet{10.1145/3522664.3528594} found that employers used this title for three distinct profiles: a data science engineer focused on modeling, an AI software engineer focused on software development, and a generalist AI engineer covering both.
The job tasks of these profiles ranged from business understanding and data engineering to modeling, software development, and operations engineering, and the authors concluded that the role was less established than that of a data scientist.
In other words, the title ``AI engineer'' did not denote one skill profile, but several, each of which overlaps with one or more of our four reference roles.
Since the rise of LLMs, the title has arguably come to cover the development of LLM-based applications, which overlaps with prompt engineering itself.
A comparison against such a heterogeneous role would not provide a stable point of reference. Any difference or similarity to prompt engineering could reflect the mix of profiles in the sample rather than the role itself.}
The machine learning engineer, as the most AI-adjacent established role in April 2025, provides a particularly conservative comparison: if prompt engineering differs from it, the distinction is unlikely to be an artifact of the baseline selection.%
}%

The data collection was conducted in a two-step process: 
	First we gathered all links to job postings that match the search criteria,
	then we extracted detailed information for each job posting.
For identifying job roles, we searched for exact matches in job titles, using quoted keywords to ensure precise matching jobs (i.e., ``prompt engineer'').
This approach allowed us to create a precise snapshot of the job market for these five job roles at a specific point in time.

\subsection{Data Preprocessing}%
\ok{%
Prior to analysis, the collected data underwent two sequential preprocessing steps: first, we removed duplicate job postings to ensure each listing was unique (Section \ref{sec:deduplication}), and we then organized the remaining entries into coherent groups based on similarity (Section \ref{sec:grouping}).
}%

\subsubsection{Deduplication,}
\label{sec:deduplication}
Some of the job postings in the data are duplicates or near-duplicates. 
For instance, international companies may create the same job postings across multiple regions. These duplicate job postings are sometimes not an exact match, but differ slightly in their content.
This makes the deduplication process non-trivial.

To address this, we calculate the similarities between job postings and determine a cut-off value in similarity to filter out duplicate job postings.
To this end, we first grouped the dataset by job roles (``data engineer'', ``prompt engineer'', etc.) to reduce the computational complexity and improve accuracy.
Then, for each group, we applied 
a three step process, consisting of vocabulary construction, TF-IDF calculation for each document, and pairwise cosine-similarity comparison.
\begin{enumerate}
	\item Vocabulary construction:
	We first tokenized all documents and built a vocabulary (i.e., a set of all unique words across the entire corpus).
	The size of the vocabulary $v$ determined the dimensionality of the TF-IDF vectors.
	\item TF-IDF vector for each document:
	Each job posting was represented as a sparse vector of length $v$.
	Each component in the vector corresponds to a word in the vocabulary.
	If a word did not appear in the document, its TF-IDF value is zero.
	\item Cosine similarity:
	Given the vector representation of documents, we computed the cosine similarity between any two job postings by comparing their TF-IDF vectors.
\end{enumerate}

Based on visual inspection of the distribution of similarity scores across all job posting pairs, we can identify a distinct secondary mode at a threshold of 0.95, 
above which job postings were considered duplicates.


\subsubsection{Grouping.}
\label{sec:grouping}
Using connected component clustering, we identified groups of similar job postings and retained only one representative from each cluster.
Connected component clustering is a graph-based method that groups items into clusters by linking pairs with similarity above a defined threshold, such that each cluster corresponds to a connected component in the resulting similarity graph.
We created an undirected graph where nodes represent job postings, and edges connect nodes with similarity threshold above 0.95.
We then identified all connected components using Depth-First Search (DFS) \citep{10.1145/362248.362272}.
A connected component is a subgraph where there exists a path between any pair of nodes, meaning all job postings in the component are transitively similar to each other.
From each connected component, we randomly selected a representative job posting for further analysis.
This process reduced our initial dataset from 34,010 raw records to 20,644 unique job postings ($\approx33\%$ reduction), including 72~prompt engineer positions.
The following section describes how we extracted skills from these job postings.

\subsection{Skill Extraction}
To address RQ2 and RQ3, we initially attempted to apply the SkillSpan model \citep{zhang2022skillspanhardsoftskill} for skill extraction.
However, this approach proved suboptimal for our specific dataset.
The SkillSpan model frequently extracted entire 
phrases as skill entities,
making it hard to identify the specific skill in our analysis. Combined with the relatively low accuracy of the model when applied to our dataset, SkillSpan was ruled out for extracting skills from the job postings.%

Instead, we followed the qualitative analysis framework proposed by \citet{zhang_redefining_2024}. This framework suggests specific techniques for designing effective LLM prompts for qualitative analysis of data.
Appendix~\ref{fig:skill-extraction-prompt} contains our prompt, which has the following key elements:%
\begin{itemize}%
	\item \textit{Background understanding}: We tell the LLM it was an ``expert skill extractor'' and explain what job postings look like.
	This helps the LLM better understand what it is analyzing.
	
	\item \textit{Clear goal}: We clearly state what we want -- to identify and sort skills from job postings into categories, such as soft skills, hard skills, and qualifications.
	Being specific improved the results.
	
	\item \textit{Step-by-step process}: We give the LLM a clear process to follow: read the whole job post, find skills, and sort them into categories. This guided approach led to more consistent results.
	
	\item \textit{Input description}: We explained the input are job postings
	
	\item \textit{Structured output}: We asked for results in a specific format (JSON). The expectations of the output are described in detail to avoid any misinterpretation. This makes it easier to collect and analyze the data consistently across all job postings.
	
	\item \textit{Expert role}: By asking the LLM to act as an expert skill extractor. This has been found to be more effective \citep{dong2024selfcollaborationcodegenerationchatgpt}
	
	\item \textit{Clean data}: The data has been cleaned of unnecessary characters and change to lower case to ensure data uniform.
	
\end{itemize}%


	
	


The extracted skills contained some variation in terminology and phrasing.
For instance, ``Python programming,'' ``Python development,'' and ``coding in Python'' all refer to the same skill.
%
To standardize the extracted information, we implemented an approach using k-means clustering to group semantically similar skills.
Each extracted skill was first mapped to a high-dimensional vector using sentence transformers \citep{reimers-2019-sentence-bert}, specifically the all-MiniLM-L6-v2 variant which builds upon the MiniLM architecture \citep{wang2020minilm}.
This model creates embeddings that capture the semantics of text phrases, meaning that skills with similar meanings are located close to each other in the embedding space.
We then applied k-means clustering to these skill vectors. K-means partitions the vector space into k clusters by minimizing the within-cluster variance, assigning each skill to the nearest cluster centroid.
To determine an appropriate number of clusters, we conducted an iterative evaluation: we varied k from 5 to 20 and manually examined the top terms in each cluster to assess semantic coherence.
Based on this evaluation, we selected 8 clusters for soft skills and 12 clusters for hard skills, as these provided the most interpretable groupings.

We visualize the embedding clusters with UMAP \citep{UMAP}, a dimensionality reduction technique that maps high-dimensional data (in our case, embeddings) to a two-dimensional space while preserving both local and global structures.
UMAP was found to have advantages over techniques such as t-SNE and Principal Component Analysis \citep{Understanding-UMAP,UMAP}.

This approach helped to draw a clearer picture of requirements of the different job roles.
To name the cluster, we retrieved the top 5 skills (by term frequency) in each cluster and manually labeled each cluster based on these skills.

\subsection{Correlation of Soft Skills and Hard Skills}
In order to understand the relationships between soft skills and hard skills, we conducted an adjusted Pearson residual analysis using contingency tables~\citep{agresti2013categorical}.
This statistical approach allowed us to identify which combinations of soft skills and hard skills appear together more or less frequently than would be expected.
To this end, we constructed contingency tables with hard skills as columns and soft skills as rows, where each cell values is adjusted Pearson residual.
The adjusted Pearson residual for cell $(i,j)$ is calculated as:
\begin{equation*}
	r_{ij} = \frac{O_{ij} - E_{ij}}{\sqrt{E_{ij}(1 - p_{i+})(1 - p_{+j})}}
\end{equation*}
where $O_{ij}$ is the observed frequency in cell $(i,j)$, $E_{ij}$ is the expected frequency under independence, $p_{i+}$ is the marginal proportion for row $i$, and $p_{+j}$ is the marginal proportion for column $j$.

Adjusted Pearson residuals follow a standard normal distribution under the null hypothesis of independence. Values with $|r_{ij}| > 2.0$ ($p < 0.05$) indicate significant deviations from independence, with positive values indicating skills that appear together more often than expected and negative values indicating skills that appear together less often than expected.

\section{Results}%
\label{sec:results}%
\subsection{Descriptive Data Overview (RQ1)}

After removing duplicates, our final dataset had 20,662 unique job postings across five job roles (prompt engineer, machine learning engineer, data scientist, data engineer, and data analyst).
Despite the high relevance of LLMs in the tech industry, prompt engineer positions made up only a small fraction of our sample in April
2025 (72~jobs, less than 0.5\%), compared to the four established data and AI roles: data engineer (8,272~jobs), data analyst
(6,782~jobs), data scientist (5,524~jobs), and machine learning engineer (1,958~jobs).


\begin{figure*}[t]
	\centering
	\begin{minipage}[t]{.8\textwidth}
		\subfloat[Soft skills\label{fig:softprompt}]{
			\centering
			\includegraphics[trim=0cm 0cm 0cm 1.35cm,clip, width=\linewidth]{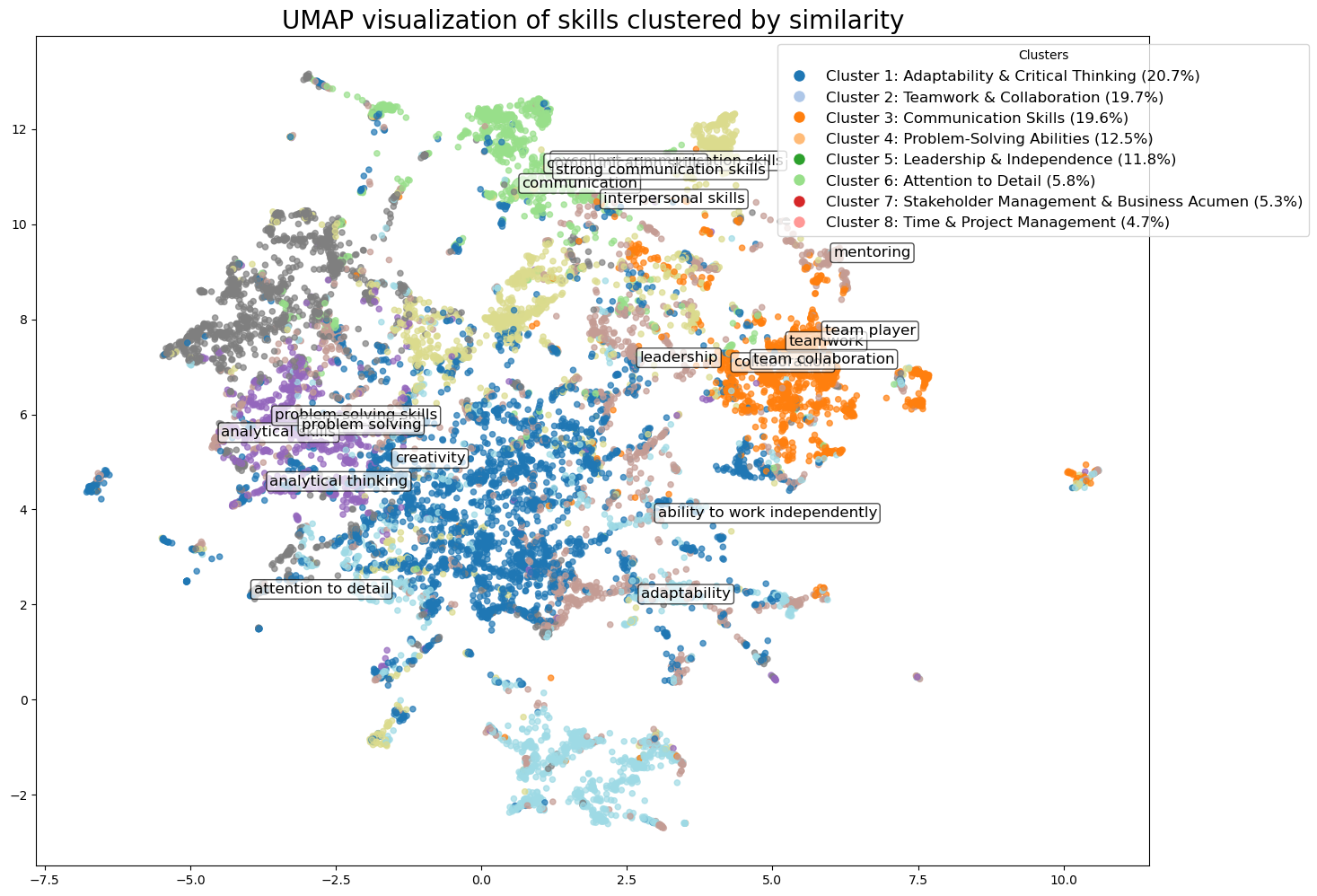}
	}
\end{minipage}
\hfill
\begin{minipage}[t]{.8\textwidth}
	\subfloat[Hard skills\label{fig:hardprompt}]{
		\centering
		\includegraphics[trim=0cm 0cm 0cm 1.4cm,clip, width=\linewidth]{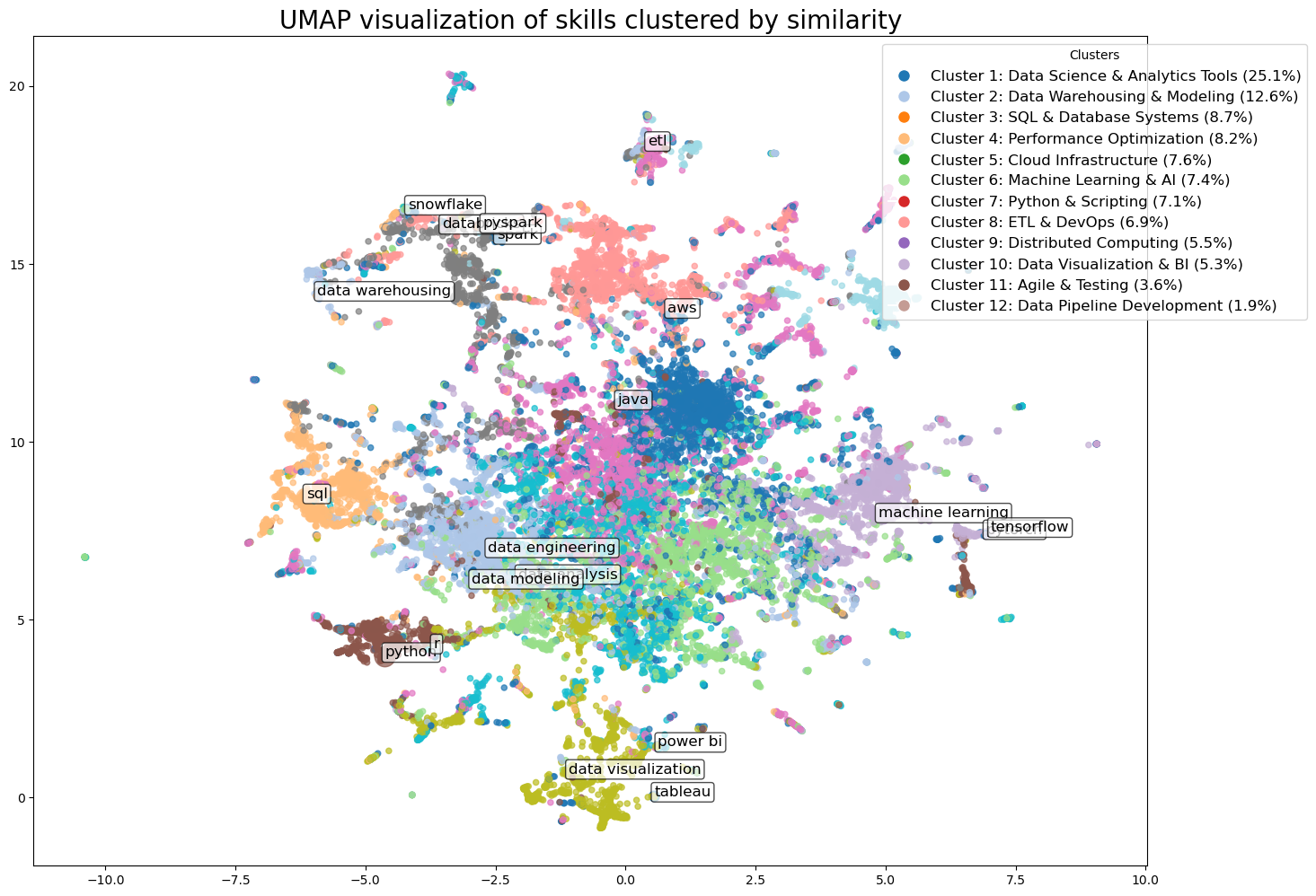}
}
\end{minipage}
\centering
\caption{Skills cluster distribution for all 
analyzed jobs. 
}
\label{fig:skills-alljobs}
\end{figure*}

This distribution confirms that while LLM use has grown quickly in industry, dedicated prompt engineer positions were still rare in April 2025, making up less than 0.5\% of the postings across the five roles in our dataset.
This matches our expectation that the emerging role of prompt engineer 
has not become as established as older positions.

\subsection{Skills Analysis Across Job Roles}
To understand what makes prompt engineering unique as a potential novel profession, we analyzed and compared both soft skills and hard skills required across the prompt engineer and the four reference roles. This analysis is contrasted with an analysis of the distinct skill profiles for prompt engineers (see Section \ref{sec:pe-skills}), giving insights into how these roles differ despite working in related areas.

\subsubsection{Soft skills requirements across jobs (RQ2).}
Figure \ref{fig:softprompt} and Table \ref{tab:soft-skills-clusters-all} show the results of our soft skill analysis across all 
collected jobs.
Our k-means clustering identified eight distinct groups of soft skills.
Statistical testing confirms significant differences between the skills profiles of the different job roles ($\chi^2(28, N=20662) = 1377.78$, $p < 0.0001$), validating that the skill requirements variation across roles is not due to random chance.


The soft skill distribution across all five roles shows that three clusters dominate the requirements: Adaptability \& Critical Thinking (20.7\%), Teamwork \& Collaboration (19.7\%), and Communication Skills (19.6\%). Together, these three groups represent about 60\% of all soft skills mentioned in job postings, showing a strong industry-wide focus on these interpersonal and thinking abilities.

Problem-Solving Abilities (12.5\%) and Leadership \& Independence (11.8\%) form the second tier of important soft skills across all five roles. The remaining clusters—Attention to Detail (5.8\%), Stakeholder Management \& Business Acumen (5.3\%), and Time \& Project Management (4.7\%)—though mentioned less often overall, show important differences between specific job types, as we'll examine in the following sections.

\subsubsection{Hard skills requirements across jobs (RQ3).}
Figure~\ref{fig:hardprompt} and Table~\ref{tab:hard-skills-clusters-all} depict the results of our comparative hard skill analysis across job postings.
We identified twelve distinct hard skill clusters that appear consistently across job postings.
Chi-square testing revealed highly significant differences among job types ($\chi^2(44, N=20662) = 67291.55$, $p<0.0001$), confirming that the technical skill profiles across different job roles represent meaningful distinctions rather than random variation.
This significant variation reflects the specialized technical requirements of each job position.
Big Data Processing (25.1\%) emerges as the most frequently required technical skill cluster across all job roles, followed by Data Warehousing \& Modeling (12.6\%) and SQL \& Database Systems (8.7\%). These three clusters form the foundation of technical requirements for data professionals, accounting for approximately 46\% of all hard skills mentioned in job postings.

\subsection{Skills Analysis for Prompt Engineers}%
\label{sec:pe-skills}%

\subsubsection{Soft skills requirements for prompt engineers (RQ2).}



\begin{table*}[!htbp]%
\centering
\input{TAB-SOFTSKILLS-PE}
\end{table*}%

\begin{table*}[!thb]
\centering
\input{TAB-SOFTSKILLS-OTHER}
\end{table*}

As depicted in Table~\ref{tab:soft-skills-clusters-all}, Figure~\ref{fig:softskills-prompt} and Table~\ref{tab:soft-skills-prompt}, prompt engineer roles place a particularly strong emphasis on collaborative communication, which together accounts for 43.8\% of their soft-skill requirements—higher than the $\approx39\%$ observed in other roles. This reflects how prompt engineers must work closely with stakeholders to translate practical needs into effective language-model prompts.

Beyond teaming and 
communication, creative problem-solving stands out: terms such as ``creative problem-solving'' and ``analytical mindset'' appear among the top skills for prompt engineers, underscoring the innovative thinking needed to design and refine prompts. The frequent mention of strong written communication further highlights the text-centric nature of the prompt engineer's work.

By contrast, time and project management skills are cited far less often in prompt engineer postings (1.9\% vs. 4.7\% across other roles), suggesting that rapid prototyping and iterative experimentation take priority over formal scheduling. When analytical abilities are required, they are most often linked with an ``ability to explain complex concepts,'' pointing to the interpretive and educational facets of the role as prompt engineers translate model behavior into actionable insights.%

\subsubsection{Hard skills requirements for prompt engineers (RQ3).}



\begin{figure*}[thbp]
\centering
\subfloat[Soft skills\label{fig:softskills-prompt}]{
\centering
\includegraphics[trim=0cm 0cm 0cm 1.35cm,clip,width=.9\linewidth]{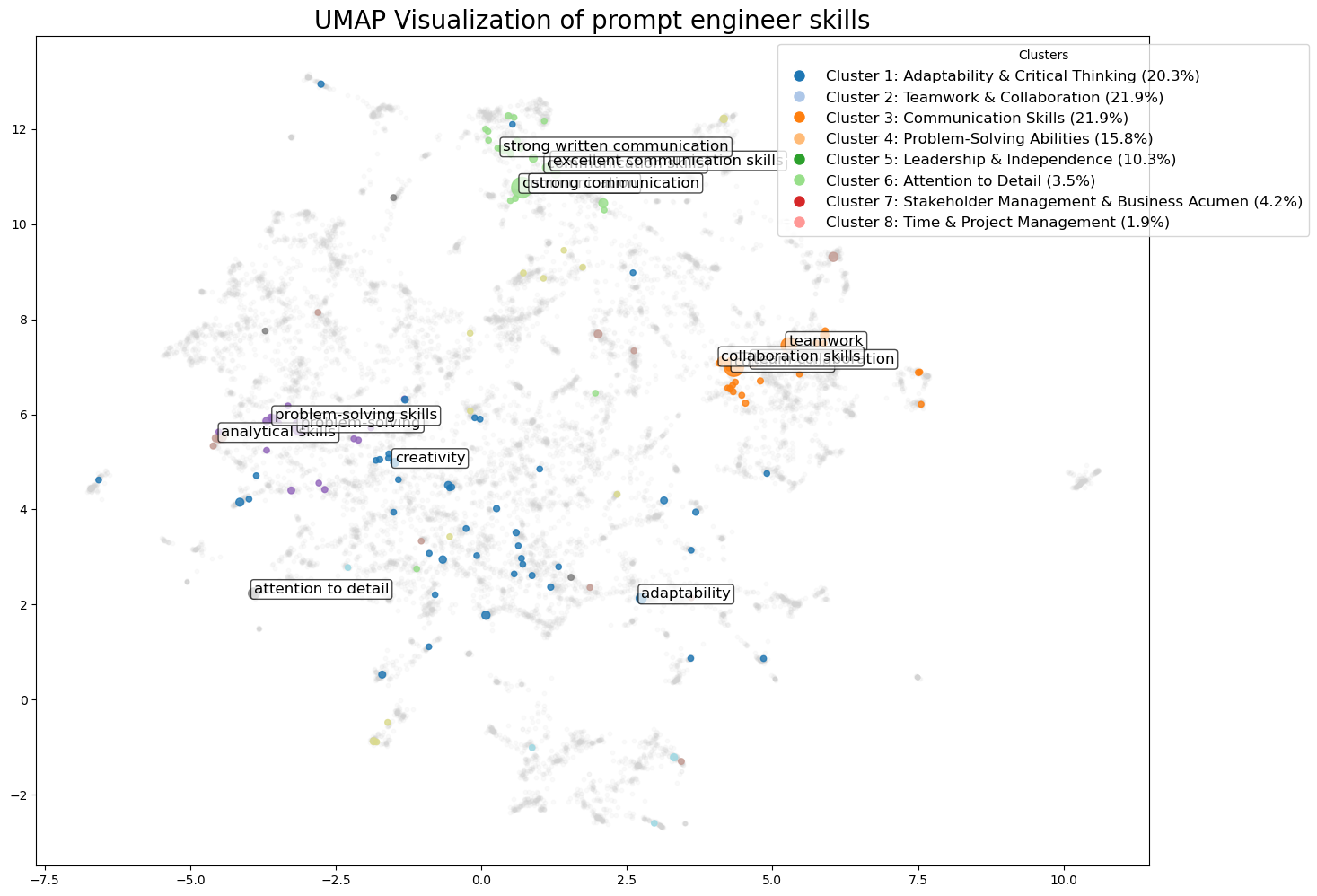}
}
\\
\subfloat[Hard skills\label{fig:hardprompt}]{
\centering
\includegraphics[trim=0cm 0cm 0cm 1.4cm,clip,width=.9\linewidth]{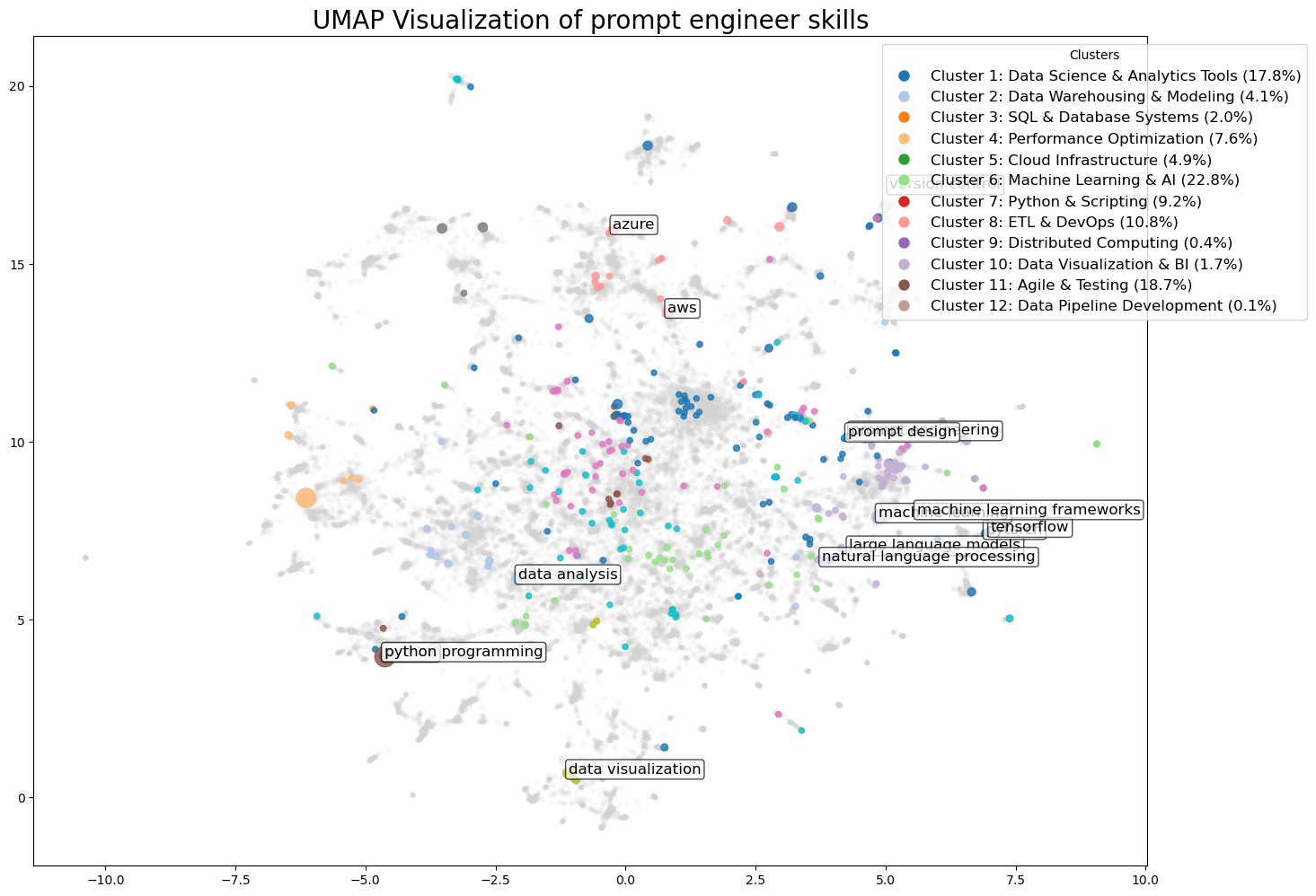}
}
\centering
\caption{Skills cluster distribution for prompt engineer positions, in comparison to skills in other jobs. Source: Authors own work.}
\label{fig:hardskills-prompt}
\end{figure*}

When examining the hard skills specifically required for prompt engineers, a distinctive technical profile emerges that highlights the unique requirements 
of this emerging role.
Figure \ref{fig:hardskills-prompt} presents the hard skills distribution specific to prompt engineer positions.
While data scientists and machine learning engineers focus on model development and statistical analysis, prompt engineers show the strongest emphasis on Machine Learning \& AI (22.8\%), specifically relating to large language models, natural language processing, and machine learning frameworks.
Also notable is the high importance of Agile \& Testing skills (18.7\%), ranking second for prompt engineers but appearing much lower for other job roles (3.6\% across all roles). This category for prompt engineers is focused on skills directly related to prompt engineering techniques (47 mentions), prompt design (6 mentions), and chain-of-thought prompting (4 mentions), highlighting the specialized nature of this role compared to traditional testing methods used by other professions.
Big Data Processing remains important (17.8\%), but with a specific focus on tools like PyTorch (10), Hugging Face (4), and LangChain (4), which are designed for working with language models rather than for traditional analytics. The importance of ETL \& DevOps (10.8\%) with emphasis on version control, Node.js, and API integration reflects the need for integrating LLM capabilities into production systems.
Python \& Scripting (9.2\%) shows the continued importance of Python (33 mentions) as the main programming language for prompt engineers, similar to other job roles, though with more specific focus on developing code that interfaces with language models.

\begin{table*}[!tbhp]
\centering
\input{TAB-HARDSKILLS-PE}
\end{table*}

\begin{table*}[!tbhp]
\centering
\input{TAB-HARDSKILLS-OTHER}
\end{table*}

Tables~\ref{tab:hard-skills-prompt} and~\ref{tab:hard-skills-clusters-all} present a side-by-side comparison of hard skill requirements across all five job roles, further illustrating the distinctive technical profile of prompt engineers. This comparison reveals several key insights:

\begin{itemize}
\item Prompt engineers require a high proportion of Machine Learning \& AI skills (22.8\%), even higher than machine learning engineers (21.6\%) and data scientists (16.0\%), yet their focus is specifically on language models rather than general machine learning algorithms.

\item Prompt engineers show a significantly higher demand for Agile \& Testing skills (18.7\%) compared to all other roles, with data analysts being the next highest at only 8.5\%. However, the specific skills within this cluster differ dramatically, with prompt engineers focusing on prompt design methods rather than traditional software testing approaches.

\item The emergence of specialized tools such as LangChain, Hugging Face, and vector databases in prompt engineer job descriptions highlights the development of a specialized toolset for this role that differs from traditional data engineering and machine learning tools.

\item Data pipeline development, which is critical for data engineers (3.1\%), is almost nonexistent for prompt engineers (0.1\%), suggesting less focus on traditional ETL workflows in the prompt engineering role and more emphasis on specialized prompt optimization processes.
\end{itemize}

The emergence of these specialized skill requirements confirms that prompt engineering 
has a distinct technical profile 
rather than simply being an extension of any one of the four reference roles.
This technical specialization, combining AI expertise with prompt-specific methods and tools, shows that prompt engineering was not simply absorbed by any of the established roles into which it might have been folded. At the same time, prompt engineer positions made up only a small segment of the postings we sampled in April 2025.%

\subsubsection{Correlation between soft and hard skills of prompt engineers.}
\label{sec:correlations}

\begin{figure*}[!thb]
\centering
\includegraphics[trim=0cm 0cm 0cm 0.7cm,clip, width=.9\linewidth]{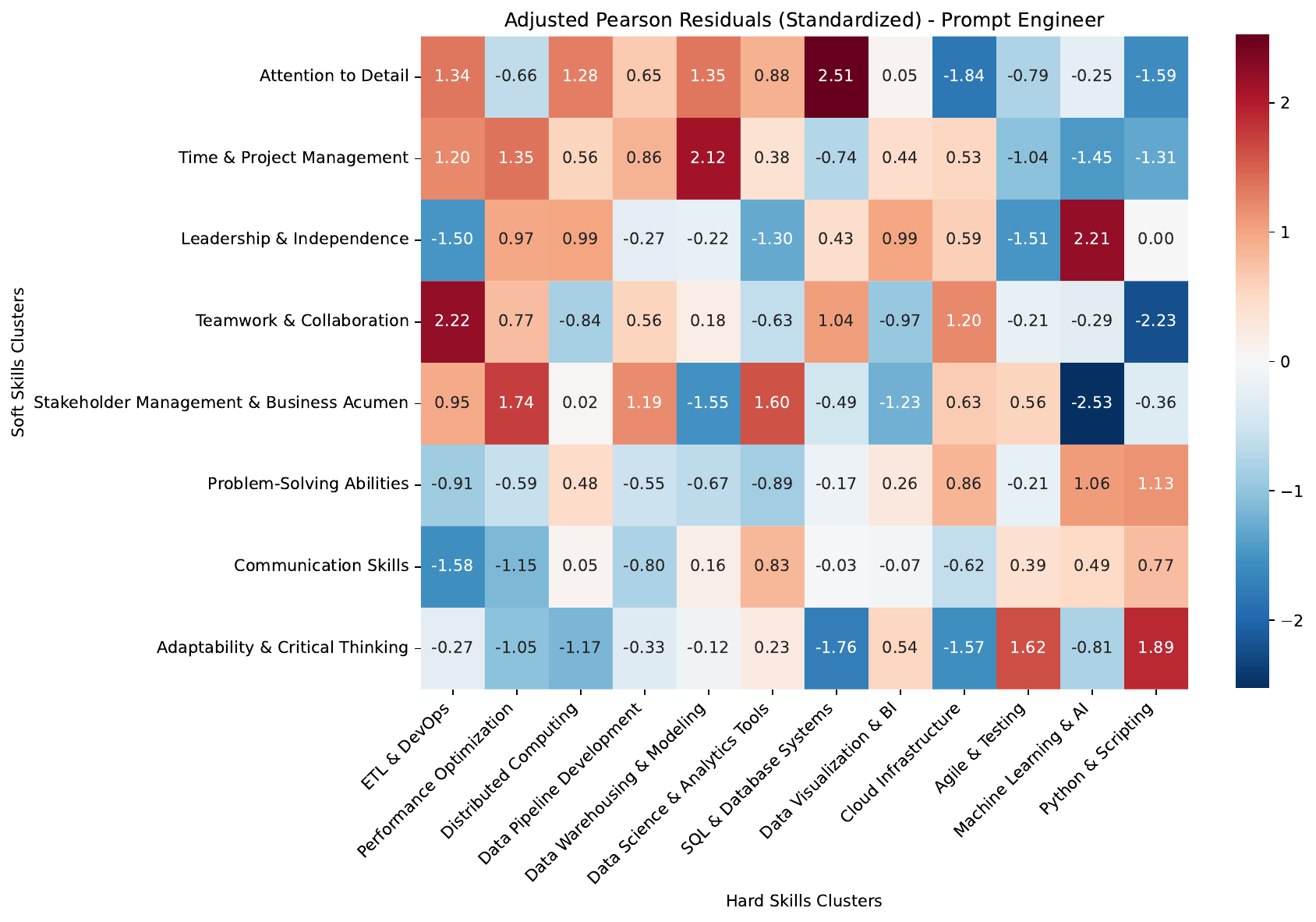}
\caption{Skills co-occurrence between soft and hard skills for prompt engineer positions.
High values indicate pairs of skills that are often mentioned together in job descriptions, whereas low values indicate the opposite.
Values of $|r| > 2.0$ indicate significant deviations from expected values ($p < 0.05$).
Source: Authors own work.
}%
\label{fig:skills-cooccurrence}%
\end{figure*}

Figure~\ref{fig:skills-cooccurrence} reveals nuanced patterns that help us understand what employers value in the role of prompt engineers beyond general skill listings.
%
The significant correlation between Teamwork \& Collaboration and ETL \& DevOps ($r = 2.22$) indicates that prompt engineers working on data integration and deployment pipelines are expected to collaborate extensively, likely because these tasks involve cross-functional workflows.
There further is a positive association between the skills Leadership \& Independence and Machine Learning (ML) \& AI ($r = 2.21$). This may suggest that high autonomy in ML-heavy prompt engineering roles is required, possibly due to the experimental and exploratory nature of prompt design for LLMs.
Further, Attention to Detail is often paired with SQL \& Database Systems skills ($r = 2.51$).
Data-related roles for prompt engineers require precision, likely for accurate data extraction or grounding prompts in structured databases.
The significant co-occurrence of Time \& Project Management and Data Warehousing \& Modeling ($r = 2.12$)
could suggest that prompt engineers working in structured data environments also manage timelines or project deliverables.
Engineers optimizing prompts for latency or token efficiency need to think critically and iterate quickly, which could explain the co-occurrence of the skills Adaptability \& Critical Thinking and Performance Optimization ($r = 1.89$).

We identified two significant negative co-occurrences of skills. 
The skills Teamwork \& Collaboration are negatively correlated with SQL \& Database Systems ($r = -2.23$).
This may imply that prompt engineers working on back-end–focused and database-oriented tasks are less integrated with team processes.
The second negative association is between Stakeholder Management \& Business Acumen and Machine Learning \& AI ($r = -2.53$).
This could indicate that prompt engineers prioritize business-facing tasks over heavy technical AI work.
It could, however, also indicate a potential technical sub-role within the prompt engineer role, where machine learning and developer-oriented tasks are emphasized over higher-level 
communication and business tasks.%

\section{Discussion}%
\label{sec:discussion}%
We humans increasingly rely on generative AI, and in particular large language models.
In our daily, but also our professional interaction with LLMs, we employ ``prompt engineering'' almost routinely to make the LLM produce our intended outcomes.
The practice and skill of prompt engineering is playing an increasingly important role in research and industry \cite{prompting-ai-art}.
In research, some have cautioned against the growing ``LLMification'' that is taking place in many research areas \citep{pang_understanding_2025}.
In industry, 
the LLM's ability for understanding natural language has unlocked capabilities and opportunities that were previously not feasible or cost-effective.
Without a doubt, technological change \citep{WEF-job-report-2025} and our increased acceptance and reliance on AI will contribute and accelerate changes in skills and jobs.
The World Economic Forum’s 2025 report on the future of jobs \citep{WEF-job-report-2025} 
highlights the close linkage of these skills with the digital transformation.
%
There will be changes in skills in the work force in the coming years, with new job roles being created \citep{nytimes}.
In this work, we examined a snapshot of the state of prompt engineering in industry, as of April 2025, and took inventory of the skills required by a potential new role of ``prompt engineer.''
In the remainder of this section, we discuss our findings on the emergence of prompt engineering as a profession in industry, its skill requirements, and implications for the job~market.%

\subsection{The Emerging Status of Prompt Engineering (RQ1)}
Our analysis shows that prompt engineering remained in its early stages
as a stand-alone profession.
With only 72~prompt engineer positions identified (less than 0.5\% of the postings across the five roles in our dataset), the role had not become widespread relative to established data and AI roles.
This limited presence suggests that, despite significant industry attention on large language models, dedicated prompt engineering positions were relatively rare.
\rev{Since our data collection, AI-related job titles, such as AI engineer and the newer AI agent engineer, have become more visible. These titles are not part of our reference frame.} Whether these roles take over responsibilities once associated with prompt engineering cannot be determined from our data.
%
This finding is notable given the timing of our snapshot. In April 2025, prompts were still written mostly by hand, which gave dedicated prompt engineering roles their strongest case. That employers nonetheless rarely created such roles suggests they treated prompt engineering as a skill within existing roles.%
\subsection{The Distinct Soft Skill Profile of Prompt Engineers (RQ2)}
While technical skills may become obsolete \citep{schumpeter.pdf}, new skills can emerge, giving rise to distinct skill profiles that are useful in a wide number of professions.
Based on our dated snapshot, prompt engineering is one such skill.
The skill profile of prompt engineers has a strong overlap with what has been termed the ``21st-century'' skill profile in prior literature \citep{VANLAAR2017577,10.3389/feduc.2024.1366434}.
For instance, van Laar et al. identified seven 21st-century skills in their systematic literature review \citep{VANLAAR2017577}: technical, information management, communication, collaboration, creativity, critical thinking, and problem solving.
In their report on the Future of Jobs, the World Economic Forum referred to these skills as ``Core skills in 2030''~\cite[p.~41]{WEF-job-report-2025}, including skills such as creative thinking, technological literacy, and AI and big data.
We find these skills are already well represented in our analysis of prompt engineering positions in April 2025.

The two key soft skills identified for the emerging role of prompt engineers are an enhanced focus on communication (21.9\%) and creative problem-solving (15.8\%).
Prompt engineer positions emphasize strong communication skills more than the other investigated job roles. This reflects the expectation that prompt engineers must effectively communicate their innovative approaches and solutions to various stakeholders.
Prompt engineers are also expected to apply creative approaches to problem-solving, with skills such as \textit{``creative mindset''} and \textit{``creative problem-solving''} appearing frequently in job postings.
This suggests employers recognize that prompting is an innovative field that requires employees to think creatively when solving problems.

Our analysis reveals that while communication and teamwork capabilities are universal soft skills across the five roles, aligning with Lievens and Sackett's findings on interpersonal skills \citep{lievens_validity_2012}, prompt engineers show distinctively high proportions of these requirements. As Table~\ref{tab:soft-skills-prompt} illustrates, Communication Skills and Teamwork \& Collaboration each represent 21.9\% of soft skill requirements, with Problem-solving Abilities (15.8\%) also emphasized more strongly compared to other job roles. Interestingly, Time \& Project Management skills appear least frequently (1.9\%), which may align with prompt engineering being an innovative emerging field that lacks standardized management processes. These findings indicate that prompt engineers serve as innovation communicators at the interface between LLM capabilities and stakeholder needs.

Our analysis also shows that the skills profile of prompt engineers is only partially reflected in existing skills taxonomies, such as the UK skills taxonomy \citep{uk-skills-taxonomy} and the Burning Glass skills taxonomy \citep{Burning_Glass_Skills_Taxonomy.pdf}.
Prompt engineers have a specific profile that 
not only leverages advanced linguistic expertise, including nuanced control of semantics, syntax, and discourse strategies, to iteratively craft and optimize prompts that effectively steer AI behavior, a dimension largely absent from traditional skills frameworks.
In addition, prompt engineers use tools, frameworks, and scripting languages, as demonstrated by our analysis of hard skills.
In the following section, we expand our discussion on the technical skills profile of prompt engineers.%
\subsection{Technical Skills Defining Prompt Engineering (RQ3)}
The analysis of hard skills provides evidence that prompt engineering is
technically distinct,
rather than simply an extension of the established roles into which it might have been folded.
Chi-square testing revealed highly significant differences among the skill profiles of different job types ($\chi^2(44, N=20662)=67291.55$, $p < 0.0001$), confirming that these technical skill profiles represent meaningful distinctions.
Notably, the side-by-side comparison with other job roles (cf. Table~\ref{tab:hard-skills-prompt}, Table~\ref{tab:hard-skills-clusters-all}, and Figure~\ref{fig:hardprompt}) highlights that prompt engineers stand out in four technical categories:
\begin{enumerate}
\item \textit{Machine Learning \& AI (22.8\%)}: According to our analysis, prompt engineers need broader AI skills than machine learning engineers (21.6\%), the most AI-adjacent role in our reference frame.
Job postings often mention large language models (22 times), natural language processing (16), and AI tools for language work.

\item \textit{Agile \& Testing (18.7\%)}: Prompt engineers need significant more testing skills than other roles. Data analysts, for instance, only require this skill in 8.5\% of the job postings.
This matches the findings by Liu and Chilton \citep{liu2023designguidelinespromptengineering}: the prompt engineering process involves extensive testing and optimization.

\item \textit{ETL \& DevOps (10.8\%)}: Skills such as version control and API integration are important. This shows companies want to connect LLMs to their existing systems and workflows. This aligns with our experience when manually labeling a small subset of the job postings in our initial exploration of the data.

\item \textit{Emerging Tools and Domain-specific Methods (18.7\%)}: Job postings often mention prompt engineering techniques (47 times), prompt design (6), and chain-of-thought prompting (4). These are special methods specific to this job role.
Requirements for specific tools, such as LangChain (4), Hugging Face (4), and vector databases (3), show a unique toolkit is forming for prompt engineers.
\end{enumerate}

In addition, prompt engineers require
\textit{Python \& Scripting (9.2\%)} skills.
While Python is often used for ETL, statistical analysis, machine learning pipelines, or data visualization in traditional job roles,
prompt engineers use Python specifically
to interface with and manipulate large language models (LLMs).
For instance, prompt engineers may use Python to orchestrate API calls to LLMs, construct dynamic prompts, parse model outputs, and evaluate generations -- tasks that are more about language manipulation and real-time interaction than batch data processing.
Prompt engineers also write lightweight scripts to prototype interactions, automate prompt variation, or integrate LLMs into applications, often with Python-based tools and libraries, such as 
LangChain or OpenAI.

Interestingly, \textit{Cloud Skills (4.9\%)} were found to be less important than for the four reference roles.
This may be because prompt engineering is often positioned closer to application-level prototyping and model interaction rather than infrastructure deployment or scalable production workflows.
Prompt engineers frequently work with hosted APIs and specific online tools (e.g., OpenAI, Anthropic, Cohere) rather than deploying models themselves, while
cloud infrastructure (e.g., AWS, GCP, Azure) is more relevant for backend developers, machine learning engineers, or MLOps roles responsible for running and scaling models, data pipelines, or inference services.
As a result, prompt engineers often operate in environments abstracted from cloud infrastructure, relying on tools that hide those complexities (e.g., notebooks, playgrounds, or SDKs).
Thus, while cloud proficiency can be useful, it is typically not a central requirement in prompt engineering job descriptions.

This unique mix of skills shows prompt engineering is
focused on making language models work well in real applications, with emerging
consequences
for the job market.

\subsection{Implications for the Job Market}
\ok{%
The emergence of prompt engineering as a distinct 
and specialized skill profile has several implications for both job seekers and employers:%
}%
\begin{enumerate}
\item \textit{Hybrid skill set requirements}:
Prompt engineers need a unique combination of technical AI knowledge,  strong communication, and strong creative skills.
This hybrid profile is 
rare, as it blends competencies from traditionally separate domains, such as software engineering, linguistics, and design. In practice, prompt engineers must be capable of understanding model behavior, crafting precise language instructions, and iteratively refining outputs based on feedback.
For employers, this means recruiting for a complex and still evolving role that few candidates are formally trained for, making talent acquisition and role definition more difficult.

\item \textit{Professional development pathways}:
The distinct skill profile of prompt engineering creates new opportunities for professional specialization, especially for individuals already working in adjacent fields.
Data scientists, NLP practitioners, UX designers, and machine learning engineers may find in prompt engineering a logical progression or pivot point, particularly as organizations expand their use of large language models.
The role offers a way to deepen one's focus on LLM applications while also contributing to product design, user experience, or applied AI research. 
Even if few dedicated roles remain, the underlying skills are likely to stay valuable within these adjacent professions.

\item \textit{Educational gaps}:
Existing academic programs in computer science and data science generally do not cover the core competencies required for prompt engineering. While students may gain exposure to machine learning and natural language processing, they typically receive little to no training in prompt design, model behavior analysis, or the human-in-the-loop techniques often used in real-world deployments.
As a result, graduates entering the workforce may lack formal training 
on LLM-specific tools, evaluation practices, and interaction patterns. Addressing this gap will likely require curriculum innovation, including interdisciplinary courses that bridge AI, language, and design.%

\item \textit{Agents and the future of prompt engineering}: Since our data collection, LLM-based agents have begun to write and
	refine prompts themselves. This reduces the need for hand-written prompts, but not for the skills our analysis identified. Specifying goals,
	evaluating model output, and communicating with stakeholders remain necessary when agents do the programming. For job seekers, this suggests
	building broader AI foundations alongside prompting skills. For employers, it suggests upskilling existing staff rather than creating
	dedicated prompt engineering roles.
\end{enumerate}%
\subsection{Limitations and Future Work}
While our study provides valuable insights into prompt engineer roles as of April 2025, we acknowledge several limitations:%
\begin{enumerate}
\item \textit{Sample size}: With only 72 prompt engineer positions identified on LinkedIn at the time the snapshot was taken, our analysis gives insights into prompt engineering as of April 2025 in an important online job market, but the result may not be indicative of wider developments.%

\item \textit{Search keywords}: We specifically searched for ``prompt engineer'' as a job role. However, some job advertisements may mention prompt engineering as a skill. These advertisements are not captured by our analysis, because our key subject of study is the role of prompt engineer as a separate job role.%

\item \textit{Data processing}: Specifically the clustering process this may still contains error because the clustering process are not perfect as the embedding is missing the context. We need an embedding that's more context aware or a different clustering/grouping method.%

\item \textit{Skills extraction}: In our work, we focused on extracting soft and hard skills. This analysis could be expanded into extracting competencies, which include skills, knowledge, attitudes, and abilities \citep{Skills-Taxonomy_Final-1.pdf}.
Such an analysis would provide a more nuanced understanding of the job profile of the prompt engineer.


\item \textit{Geographic limitations}: Our data collection focused on LinkedIn job postings in countries with technological expertise, potentially missing regional variations in how prompt engineering roles are defined.%

\item \textit{Temporal dynamics}: Given the rapid evolution of LLM technology and methods, the skill requirements for prompt engineers are likely to change quickly, requiring regular update.%

\item \textit{Choice of comparison roles}: Our reference frame consists of four established data and AI roles.
Roles that existed before, but gained strong prominence after our data collection, such as AI engineer \cite{10.1145/3522664.3528594}, are not represented. Future work could examine whether, and to what extent, the prompt engineering skills identified here appear in the requirements of these roles.%
\end{enumerate}

Future research could extend the analysis to a broader dataset, and also track how prompt engineering job descriptions evolve over time as the field matures, in relation to other job descriptions.
Further, as mentioned in Section \ref{sec:correlations}, there could be distinct prompt engineering sub-roles focused more on technical or business aspects. Future work could investigate this further.
On the other hand, it is likely that prompt engineering dissolves into other roles, so future studies could examine how skills and expertise related to prompt engineering appears in job descriptions.
Longitudinal data collection would enable the study of how new professions evolve.
Additionally, qualitative studies involving interviews with  prompt engineers and their employers would provide deeper insights into the day-to-day responsibilities and challenges of this role. Finally, expanding the analysis to include salary data would help assess how the market values these specialized skills compared to traditional job roles.%
\section{Conclusion}%
\label{sec:conclusion}%
Our data-driven analysis of job postings reveals the dedicated role of prompt engineer remained rare, comprising less than 0.5\% of the postings across the five roles we sampled despite growing LLM adoption. 
Prompt engineering roles show a distinctive skill profile combining strong communication and creative problem-solving with specialized AI knowledge, which sets them apart from all four established data and AI roles into which prompt engineering might otherwise have been folded.
Most surprisingly, prompt engineering extends far beyond crafting textual prompts. The emphasis on testing and DevOps skills indicates that prompt engineers are professionals responsible for the entire solution lifecycle: designing, testing, integrating, and deploying LLM systems into production workflows.
This creates both opportunities and challenges, with important caveats about the job role's future. The combination of slow organizational adoption and rapid AI evolution creates uncertainty about the longevity of dedicated prompt engineering positions. 
Since our data collection, LLM-based agents have taken over parts of prompt writing, and we expect fewer dedicated prompt engineering positions as a result.
Job seekers should develop broader AI foundations alongside prompt engineering skills, while employers may favor upskilling existing employees over creating specialized roles.

\bibliographystyle{ACM-Reference-Format}
\bibliography{main}

\appendix
\section{Prompt for Skill Extraction from Job Postings}

{%
	\footnotesize%
	\texttt{%
		\raggedright%
		\input{PROMPT}%
	}%
	\label{fig:skill-extraction-prompt}%
}%

\end{document}

%% file: TAB-SOFTSKILLS-PE.tex
\caption{Soft skills clusters in prompt engineer postings (n = 72). Bars: share of each cluster among all soft-skill mentions extracted for a job role (\%).}
\label{tab:soft-skills-prompt}
\centering
\footnotesize
\begin{tabularx}{\textwidth}{
    p{2.6cm}
    X
    c
    l
}
\toprule
\textbf{Cluster} & \textbf{Top Skills (count)} & \textbf{Freq.} & \input{bars/soft-axis} \\
\midrule
Teamwork \& Collaboration & 
Collaboration (27), Teamwork (10), Team collaboration (5), Collaboration skills (5), Team player (3)
 & 21.9\% & \input{bars/soft-teamwork-collaboration} \\
\midrule
Communi\-cation Skills & 
Communication (15), Communication skills (10), Excellent communication skills (4), Strong written communication (4), Excellent communication (3)
 & 21.9\% & \input{bars/soft-communication-skills} \\
\midrule
Adaptability \& Critical Thinking & 
Creativity (7), Adaptability (6), Self-driven (2), Creative mindset (2), Innovation (2)
 & 20.3\% & \input{bars/soft-adaptability-critical-thinking} \\
\midrule
Problem-Solving Abilities & 
Problem-solving (14), Problem-solving skills (13), Problem-solving mindset (3), Analytical mindset (3), Creative problem-solving (3)
 & 15.8\% & \input{bars/soft-problem-solving-abilities} \\
\midrule
Leadership \& Independence & 
Analytical skills (16), Strong analytical skills (3), Project management skills (2), Organizational skills (2), Ability to explain complex concepts (2)
 & 10.3\% & \input{bars/soft-leadership-independence} \\
\midrule
Stakeholder Management \& Business Acumen & 
Ownership (2), Business judgement (1), Consultative culture (1), Client collaboration (1), Ownership and accountability (1)
 & 4.2\% & \input{bars/soft-stakeholder-management-business-acumen} \\
\midrule
Attention to Detail & 
Attention to detail (7), Documentation (1), Exceptional attention to detail (1), Quick learner (1), Ability to identify patterns (1)
 & 3.5\% & \input{bars/soft-attention-to-detail} \\
\midrule
Time \& Project Management & 
Project management (3), Excellent time management (1), Ability to prioritize workload (1), Fast execution mindset (1)
 & 1.9\% & \input{bars/soft-time-project-management} \\
\bottomrule
\end{tabularx}

%% file: bars/soft-axis.tex
\begin{tikzpicture}[x=0.08667cm,baseline=(lbl.base),font=\fontsize{5}{5.5}\selectfont]
\node[anchor=east,inner sep=1.5pt] (lbl) at (0,0) {\phantom{Prompt engineer}};
\draw[black!70,line width=0.4pt] (0,0) -- (30,0);
\draw[black!70,line width=0.4pt] (0,0) -- (0,2pt) node[above,inner sep=0.5pt] {0};
\draw[black!70,line width=0.4pt] (10,0) -- (10,2pt) node[above,inner sep=0.5pt] {10};
\draw[black!70,line width=0.4pt] (20,0) -- (20,2pt) node[above,inner sep=0.5pt] {20};
\draw[black!70,line width=0.4pt] (30,0) -- (30,2pt) node[above,inner sep=0.5pt] {30};
\node[anchor=west,inner sep=1pt,xshift=5pt] at (30,0.9ex) {\%};
\end{tikzpicture}%

%% file: bars/soft-teamwork-collaboration.tex
\begin{tikzpicture}[x=0.08667cm,y=-5.60pt,baseline={([yshift=-1.2ex]current bounding box.north)},font=\fontsize{5}{5.5}\selectfont]
\draw[black!18,line width=0.3pt] (10,-0.5) -- (10,4.5);
\draw[black!18,line width=0.3pt] (20,-0.5) -- (20,4.5);
\draw[black!18,line width=0.3pt] (30,-0.5) -- (30,4.5);
\fill[fill={rgb,255:red,148;green,103;blue,189}] (0,-0.31) rectangle (21.930,0.31);
\node[anchor=east,inner sep=1.5pt] at (0,0) {Prompt engineer};
\fill[fill={rgb,255:red,214;green,39;blue,40}] (0,0.69) rectangle (24.567,1.31);
\node[anchor=east,inner sep=1.5pt] at (0,1) {ML engineer};
\fill[fill={rgb,255:red,44;green,160;blue,44}] (0,1.69) rectangle (19.472,2.31);
\node[anchor=east,inner sep=1.5pt] at (0,2) {Data scientist};
\fill[fill={rgb,255:red,255;green,127;blue,14}] (0,2.69) rectangle (21.853,3.31);
\node[anchor=east,inner sep=1.5pt] at (0,3) {Data engineer};
\fill[fill={rgb,255:red,31;green,119;blue,180}] (0,3.69) rectangle (15.851,4.31);
\node[anchor=east,inner sep=1.5pt] at (0,4) {Data analyst};
\draw[black!70,line width=0.4pt] (0,-0.5) -- (0,4.5);
\end{tikzpicture}%

%% file: bars/soft-communication-skills.tex
\begin{tikzpicture}[x=0.08667cm,y=-5.60pt,baseline={([yshift=-1.2ex]current bounding box.north)},font=\fontsize{5}{5.5}\selectfont]
\draw[black!18,line width=0.3pt] (10,-0.5) -- (10,4.5);
\draw[black!18,line width=0.3pt] (20,-0.5) -- (20,4.5);
\draw[black!18,line width=0.3pt] (30,-0.5) -- (30,4.5);
\fill[fill={rgb,255:red,148;green,103;blue,189}] (0,-0.31) rectangle (21.930,0.31);
\node[anchor=east,inner sep=1.5pt] at (0,0) {Prompt engineer};
\fill[fill={rgb,255:red,214;green,39;blue,40}] (0,0.69) rectangle (18.386,1.31);
\node[anchor=east,inner sep=1.5pt] at (0,1) {ML engineer};
\fill[fill={rgb,255:red,44;green,160;blue,44}] (0,1.69) rectangle (20.663,2.31);
\node[anchor=east,inner sep=1.5pt] at (0,2) {Data scientist};
\fill[fill={rgb,255:red,255;green,127;blue,14}] (0,2.69) rectangle (18.697,3.31);
\node[anchor=east,inner sep=1.5pt] at (0,3) {Data engineer};
\fill[fill={rgb,255:red,31;green,119;blue,180}] (0,3.69) rectangle (20.171,4.31);
\node[anchor=east,inner sep=1.5pt] at (0,4) {Data analyst};
\draw[black!70,line width=0.4pt] (0,-0.5) -- (0,4.5);
\end{tikzpicture}%

%% file: bars/soft-adaptability-critical-thinking.tex
\begin{tikzpicture}[x=0.08667cm,y=-5.60pt,baseline={([yshift=-1.2ex]current bounding box.north)},font=\fontsize{5}{5.5}\selectfont]
\draw[black!18,line width=0.3pt] (10,-0.5) -- (10,4.5);
\draw[black!18,line width=0.3pt] (20,-0.5) -- (20,4.5);
\draw[black!18,line width=0.3pt] (30,-0.5) -- (30,4.5);
\fill[fill={rgb,255:red,148;green,103;blue,189}] (0,-0.31) rectangle (20.327,0.31);
\node[anchor=east,inner sep=1.5pt] at (0,0) {Prompt engineer};
\fill[fill={rgb,255:red,214;green,39;blue,40}] (0,0.69) rectangle (23.741,1.31);
\node[anchor=east,inner sep=1.5pt] at (0,1) {ML engineer};
\fill[fill={rgb,255:red,44;green,160;blue,44}] (0,1.69) rectangle (21.565,2.31);
\node[anchor=east,inner sep=1.5pt] at (0,2) {Data scientist};
\fill[fill={rgb,255:red,255;green,127;blue,14}] (0,2.69) rectangle (20.068,3.31);
\node[anchor=east,inner sep=1.5pt] at (0,3) {Data engineer};
\fill[fill={rgb,255:red,31;green,119;blue,180}] (0,3.69) rectangle (19.863,4.31);
\node[anchor=east,inner sep=1.5pt] at (0,4) {Data analyst};
\draw[black!70,line width=0.4pt] (0,-0.5) -- (0,4.5);
\end{tikzpicture}%

%% file: bars/soft-problem-solving-abilities.tex
\begin{tikzpicture}[x=0.08667cm,y=-5.60pt,baseline={([yshift=-1.2ex]current bounding box.north)},font=\fontsize{5}{5.5}\selectfont]
\draw[black!18,line width=0.3pt] (10,-0.5) -- (10,4.5);
\draw[black!18,line width=0.3pt] (20,-0.5) -- (20,4.5);
\draw[black!18,line width=0.3pt] (30,-0.5) -- (30,4.5);
\fill[fill={rgb,255:red,148;green,103;blue,189}] (0,-0.31) rectangle (15.800,0.31);
\node[anchor=east,inner sep=1.5pt] at (0,0) {Prompt engineer};
\fill[fill={rgb,255:red,214;green,39;blue,40}] (0,0.69) rectangle (13.522,1.31);
\node[anchor=east,inner sep=1.5pt] at (0,1) {ML engineer};
\fill[fill={rgb,255:red,44;green,160;blue,44}] (0,1.69) rectangle (11.970,2.31);
\node[anchor=east,inner sep=1.5pt] at (0,2) {Data scientist};
\fill[fill={rgb,255:red,255;green,127;blue,14}] (0,2.69) rectangle (13.205,3.31);
\node[anchor=east,inner sep=1.5pt] at (0,3) {Data engineer};
\fill[fill={rgb,255:red,31;green,119;blue,180}] (0,3.69) rectangle (11.584,4.31);
\node[anchor=east,inner sep=1.5pt] at (0,4) {Data analyst};
\draw[black!70,line width=0.4pt] (0,-0.5) -- (0,4.5);
\end{tikzpicture}%

%% file: bars/soft-leadership-independence.tex
\begin{tikzpicture}[x=0.08667cm,y=-5.60pt,baseline={([yshift=-1.2ex]current bounding box.north)},font=\fontsize{5}{5.5}\selectfont]
\draw[black!18,line width=0.3pt] (10,-0.5) -- (10,4.5);
\draw[black!18,line width=0.3pt] (20,-0.5) -- (20,4.5);
\draw[black!18,line width=0.3pt] (30,-0.5) -- (30,4.5);
\fill[fill={rgb,255:red,148;green,103;blue,189}] (0,-0.31) rectangle (10.323,0.31);
\node[anchor=east,inner sep=1.5pt] at (0,0) {Prompt engineer};
\fill[fill={rgb,255:red,214;green,39;blue,40}] (0,0.69) rectangle (10.340,1.31);
\node[anchor=east,inner sep=1.5pt] at (0,1) {ML engineer};
\fill[fill={rgb,255:red,44;green,160;blue,44}] (0,1.69) rectangle (11.375,2.31);
\node[anchor=east,inner sep=1.5pt] at (0,2) {Data scientist};
\fill[fill={rgb,255:red,255;green,127;blue,14}] (0,2.69) rectangle (11.999,3.31);
\node[anchor=east,inner sep=1.5pt] at (0,3) {Data engineer};
\fill[fill={rgb,255:red,31;green,119;blue,180}] (0,3.69) rectangle (12.435,4.31);
\node[anchor=east,inner sep=1.5pt] at (0,4) {Data analyst};
\draw[black!70,line width=0.4pt] (0,-0.5) -- (0,4.5);
\end{tikzpicture}%

%% file: bars/soft-stakeholder-management-business-acumen.tex
\begin{tikzpicture}[x=0.08667cm,y=-5.60pt,baseline={([yshift=-1.2ex]current bounding box.north)},font=\fontsize{5}{5.5}\selectfont]
\draw[black!18,line width=0.3pt] (10,-0.5) -- (10,4.5);
\draw[black!18,line width=0.3pt] (20,-0.5) -- (20,4.5);
\draw[black!18,line width=0.3pt] (30,-0.5) -- (30,4.5);
\fill[fill={rgb,255:red,148;green,103;blue,189}] (0,-0.31) rectangle (4.182,0.31);
\node[anchor=east,inner sep=1.5pt] at (0,0) {Prompt engineer};
\fill[fill={rgb,255:red,214;green,39;blue,40}] (0,0.69) rectangle (3.874,1.31);
\node[anchor=east,inner sep=1.5pt] at (0,1) {ML engineer};
\fill[fill={rgb,255:red,44;green,160;blue,44}] (0,1.69) rectangle (5.631,2.31);
\node[anchor=east,inner sep=1.5pt] at (0,2) {Data scientist};
\fill[fill={rgb,255:red,255;green,127;blue,14}] (0,2.69) rectangle (5.139,3.31);
\node[anchor=east,inner sep=1.5pt] at (0,3) {Data engineer};
\fill[fill={rgb,255:red,31;green,119;blue,180}] (0,3.69) rectangle (5.631,4.31);
\node[anchor=east,inner sep=1.5pt] at (0,4) {Data analyst};
\draw[black!70,line width=0.4pt] (0,-0.5) -- (0,4.5);
\end{tikzpicture}%

%% file: bars/soft-attention-to-detail.tex
\begin{tikzpicture}[x=0.08667cm,y=-5.60pt,baseline={([yshift=-1.2ex]current bounding box.north)},font=\fontsize{5}{5.5}\selectfont]
\draw[black!18,line width=0.3pt] (10,-0.5) -- (10,4.5);
\draw[black!18,line width=0.3pt] (20,-0.5) -- (20,4.5);
\draw[black!18,line width=0.3pt] (30,-0.5) -- (30,4.5);
\fill[fill={rgb,255:red,148;green,103;blue,189}] (0,-0.31) rectangle (3.535,0.31);
\node[anchor=east,inner sep=1.5pt] at (0,0) {Prompt engineer};
\fill[fill={rgb,255:red,214;green,39;blue,40}] (0,0.69) rectangle (2.988,1.31);
\node[anchor=east,inner sep=1.5pt] at (0,1) {ML engineer};
\fill[fill={rgb,255:red,44;green,160;blue,44}] (0,1.69) rectangle (5.372,2.31);
\node[anchor=east,inner sep=1.5pt] at (0,2) {Data scientist};
\fill[fill={rgb,255:red,255;green,127;blue,14}] (0,2.69) rectangle (4.696,3.31);
\node[anchor=east,inner sep=1.5pt] at (0,3) {Data engineer};
\fill[fill={rgb,255:red,31;green,119;blue,180}] (0,3.69) rectangle (8.266,4.31);
\node[anchor=east,inner sep=1.5pt] at (0,4) {Data analyst};
\draw[black!70,line width=0.4pt] (0,-0.5) -- (0,4.5);
\end{tikzpicture}%

%% file: bars/soft-time-project-management.tex
\begin{tikzpicture}[x=0.08667cm,y=-5.60pt,baseline={([yshift=-1.2ex]current bounding box.north)},font=\fontsize{5}{5.5}\selectfont]
\draw[black!18,line width=0.3pt] (10,-0.5) -- (10,4.5);
\draw[black!18,line width=0.3pt] (20,-0.5) -- (20,4.5);
\draw[black!18,line width=0.3pt] (30,-0.5) -- (30,4.5);
\fill[fill={rgb,255:red,148;green,103;blue,189}] (0,-0.31) rectangle (1.934,0.31);
\node[anchor=east,inner sep=1.5pt] at (0,0) {Prompt engineer};
\fill[fill={rgb,255:red,214;green,39;blue,40}] (0,0.69) rectangle (2.581,1.31);
\node[anchor=east,inner sep=1.5pt] at (0,1) {ML engineer};
\fill[fill={rgb,255:red,44;green,160;blue,44}] (0,1.69) rectangle (3.950,2.31);
\node[anchor=east,inner sep=1.5pt] at (0,2) {Data scientist};
\fill[fill={rgb,255:red,255;green,127;blue,14}] (0,2.69) rectangle (4.322,3.31);
\node[anchor=east,inner sep=1.5pt] at (0,3) {Data engineer};
\fill[fill={rgb,255:red,31;green,119;blue,180}] (0,3.69) rectangle (6.200,4.31);
\node[anchor=east,inner sep=1.5pt] at (0,4) {Data analyst};
\draw[black!70,line width=0.4pt] (0,-0.5) -- (0,4.5);
\end{tikzpicture}%

%% file: TAB-SOFTSKILLS-OTHER.tex
\caption{Soft skills clusters in the postings of the four reference roles (pooled).}
\label{tab:soft-skills-clusters-all}
\centering
\footnotesize
\begin{tabularx}{\textwidth}{
    p{3.2cm}
    X
    c
}
\toprule
\textbf{Cluster} & \textbf{Top Skills (count)} 
    & \textbf{Freq.} \\
\midrule
Adaptability \& Critical Thinking & 
adaptability (1509), creativity (757), curiosity (722), critical thinking (666), self-motivated (476)
& 20.7\% \\
Teamwork \& Collaboration & 
collaboration (7462), teamwork (4225), team collaboration (1066), team player (796), collaborative (444)
& 19.7\% \\
Communi\-cation Skills & 
communication (8114), excellent communication skills (1163), communication skills (987), strong communication skills (986), interpersonal skills (961)
& 19.6\% \\
Problem-Solving Abilities & 
problem-solving (5053), problem-solving skills (1117), problem solving (974), analytical thinking (775), strong problem-solving skills (477)
& 12.5\% \\
Leadership \& Independence & 
analytical skills (1672), mentoring (1068), leadership (1042), ability to work independently (782), mentorship (665)
& 11.8\% \\
Attention to Detail & 
attention to detail (1883), detail-oriented (450), strong attention to detail (148), results-oriented (89), detail oriented (72)
& 5.8\% \\
Stakeholder Management \& Business Acumen & 
stakeholder management (459), ownership (365), relationship building (247), stakeholder engagement (229), business acumen (165)
& 5.3\% \\
Time \& Project Management & 
time management (397), project management (391), prioritization (99), multi-tasking (95), organized (87)
& 4.7\% \\
\bottomrule
\end{tabularx}

%% file: TAB-HARDSKILLS-PE.tex
\caption{Hard skills 
	in prompt engineer postings (n = 72). Bars: share of each cluster among all hard-skill mentions extracted for a job role (\%).}
\label{tab:hard-skills-prompt}
\centering
\footnotesize
\begin{tabularx}{\textwidth}{
    p{2.6cm}
    X
    c
    l
}
\toprule
\textbf{Cluster} & \textbf{Top Skills (count)} & \textbf{Freq.} & \input{bars/hard-axis} \\
\midrule
Machine Learning \& AI & 
Large language models (22), Natural language processing (16), Tensorflow (10), Machine learning (9), Machine learning frameworks (5)
 & 22.8\% & \input{bars/hard-machine-learning-ai} \\
\midrule
Agile \& Testing & 
Prompt engineering (47), Prompt design (6), Chain-of-thought prompting (4), A/B testing (3), Writing prompts (3)
 & 18.7\% & \input{bars/hard-agile-testing} \\
\midrule
Big Data Processing & 
Pytorch (10), Typescript (4), Hugging Face (4), Langchain (4), NLP (4)
 & 17.8\% & \input{bars/hard-data-science-analytics-tools} \\
\midrule
ETL \& Dev\-Ops & 
Version control (4), Node.js (4), API integration (3), Chatbot development (3), Automation tools (3)
 & 10.8\% & \input{bars/hard-etl-devops} \\
\midrule
Python \& Scripting & 
Python (33), Python programming (7), Programming languages (4), Python development (3), Scripting (3)
 & 9.2\% & \input{bars/hard-python-scripting} \\
\midrule
Performance Optimization & 
Retrieval-augmented generation (4), Statistical modeling (2), Instruction tuning (2), Scalability (2), Optimizing prompts (2)
 & 7.6\% & \input{bars/hard-performance-optimization} \\
\midrule
Cloud Infrastructure & 
Azure (5), AWS (4), Cloud platforms (4), Microsoft Azure (3), Cloud-based application development (2)
 & 4.9\% & \input{bars/hard-cloud-infrastructure} \\
\midrule
Data Warehousing \& Modeling & 
Data analysis (14), Data science (2), Data manipulation (2), Dataset creation (1), Data transformation (1)
 & 4.1\% & \input{bars/hard-data-warehousing-modeling} \\
\midrule
SQL \& Database Systems & 
SQL (4), Vector databases (3), Database management (2), Vector DB (1), Postgresql (1)
 & 2.0\% & \input{bars/hard-sql-database-systems} \\
\midrule
Data Visualization \& BI & 
Data visualization (6), Data visualization tools (3), Google analytics (1), Analytics (1), Visualization tools (1)
 & 1.7\% & \input{bars/hard-data-visualization-bi} \\
\midrule
Distributed Computing & 
Big data tools (1), Pyspark (1), Databricks (1)
 & 0.4\% & \input{bars/hard-distributed-computing} \\
\midrule
Data Pipeline Development & 
RAG pipelines (1)
 & 0.1\% & \input{bars/hard-data-pipeline-development} \\
\bottomrule
\end{tabularx}%

%% file: bars/hard-axis.tex
\begin{tikzpicture}[x=0.08667cm,baseline=(lbl.base),font=\fontsize{5}{5.5}\selectfont]
\node[anchor=east,inner sep=1.5pt] (lbl) at (0,0) {\phantom{Prompt engineer}};
\draw[black!70,line width=0.4pt] (0,0) -- (30,0);
\draw[black!70,line width=0.4pt] (0,0) -- (0,2pt) node[above,inner sep=0.5pt] {0};
\draw[black!70,line width=0.4pt] (10,0) -- (10,2pt) node[above,inner sep=0.5pt] {10};
\draw[black!70,line width=0.4pt] (20,0) -- (20,2pt) node[above,inner sep=0.5pt] {20};
\draw[black!70,line width=0.4pt] (30,0) -- (30,2pt) node[above,inner sep=0.5pt] {30};
\node[anchor=west,inner sep=1pt,xshift=5pt] at (30,0.9ex) {\%};
\end{tikzpicture}%

%% file: bars/hard-machine-learning-ai.tex
\begin{tikzpicture}[x=0.08667cm,y=-5.60pt,baseline={([yshift=-1.2ex]current bounding box.north)},font=\fontsize{5}{5.5}\selectfont]
\draw[black!18,line width=0.3pt] (10,-0.5) -- (10,4.5);
\draw[black!18,line width=0.3pt] (20,-0.5) -- (20,4.5);
\draw[black!18,line width=0.3pt] (30,-0.5) -- (30,4.5);
\fill[fill={rgb,255:red,148;green,103;blue,189}] (0,-0.31) rectangle (22.791,0.31);
\node[anchor=east,inner sep=1.5pt] at (0,0) {Prompt engineer};
\fill[fill={rgb,255:red,214;green,39;blue,40}] (0,0.69) rectangle (21.543,1.31);
\node[anchor=east,inner sep=1.5pt] at (0,1) {ML engineer};
\fill[fill={rgb,255:red,44;green,160;blue,44}] (0,1.69) rectangle (16.015,2.31);
\node[anchor=east,inner sep=1.5pt] at (0,2) {Data scientist};
\fill[fill={rgb,255:red,255;green,127;blue,14}] (0,2.69) rectangle (1.331,3.31);
\node[anchor=east,inner sep=1.5pt] at (0,3) {Data engineer};
\fill[fill={rgb,255:red,31;green,119;blue,180}] (0,3.69) rectangle (1.829,4.31);
\node[anchor=east,inner sep=1.5pt] at (0,4) {Data analyst};
\draw[black!70,line width=0.4pt] (0,-0.5) -- (0,4.5);
\end{tikzpicture}%

%% file: bars/hard-agile-testing.tex
\begin{tikzpicture}[x=0.08667cm,y=-5.60pt,baseline={([yshift=-1.2ex]current bounding box.north)},font=\fontsize{5}{5.5}\selectfont]
\draw[black!18,line width=0.3pt] (10,-0.5) -- (10,4.5);
\draw[black!18,line width=0.3pt] (20,-0.5) -- (20,4.5);
\draw[black!18,line width=0.3pt] (30,-0.5) -- (30,4.5);
\fill[fill={rgb,255:red,148;green,103;blue,189}] (0,-0.31) rectangle (18.735,0.31);
\node[anchor=east,inner sep=1.5pt] at (0,0) {Prompt engineer};
\fill[fill={rgb,255:red,214;green,39;blue,40}] (0,0.69) rectangle (2.209,1.31);
\node[anchor=east,inner sep=1.5pt] at (0,1) {ML engineer};
\fill[fill={rgb,255:red,44;green,160;blue,44}] (0,1.69) rectangle (2.958,2.31);
\node[anchor=east,inner sep=1.5pt] at (0,2) {Data scientist};
\fill[fill={rgb,255:red,255;green,127;blue,14}] (0,2.69) rectangle (2.008,3.31);
\node[anchor=east,inner sep=1.5pt] at (0,3) {Data engineer};
\fill[fill={rgb,255:red,31;green,119;blue,180}] (0,3.69) rectangle (8.520,4.31);
\node[anchor=east,inner sep=1.5pt] at (0,4) {Data analyst};
\draw[black!70,line width=0.4pt] (0,-0.5) -- (0,4.5);
\end{tikzpicture}%

%% file: bars/hard-data-science-analytics-tools.tex
\begin{tikzpicture}[x=0.08667cm,y=-5.60pt,baseline={([yshift=-1.2ex]current bounding box.north)},font=\fontsize{5}{5.5}\selectfont]
\draw[black!18,line width=0.3pt] (10,-0.5) -- (10,4.5);
\draw[black!18,line width=0.3pt] (20,-0.5) -- (20,4.5);
\draw[black!18,line width=0.3pt] (30,-0.5) -- (30,4.5);
\fill[fill={rgb,255:red,148;green,103;blue,189}] (0,-0.31) rectangle (17.772,0.31);
\node[anchor=east,inner sep=1.5pt] at (0,0) {Prompt engineer};
\fill[fill={rgb,255:red,214;green,39;blue,40}] (0,0.69) rectangle (26.941,1.31);
\node[anchor=east,inner sep=1.5pt] at (0,1) {ML engineer};
\fill[fill={rgb,255:red,44;green,160;blue,44}] (0,1.69) rectangle (22.441,2.31);
\node[anchor=east,inner sep=1.5pt] at (0,2) {Data scientist};
\fill[fill={rgb,255:red,255;green,127;blue,14}] (0,2.69) rectangle (27.725,3.31);
\node[anchor=east,inner sep=1.5pt] at (0,3) {Data engineer};
\fill[fill={rgb,255:red,31;green,119;blue,180}] (0,3.69) rectangle (21.809,4.31);
\node[anchor=east,inner sep=1.5pt] at (0,4) {Data analyst};
\draw[black!70,line width=0.4pt] (0,-0.5) -- (0,4.5);
\end{tikzpicture}%

%% file: bars/hard-etl-devops.tex
\begin{tikzpicture}[x=0.08667cm,y=-5.60pt,baseline={([yshift=-1.2ex]current bounding box.north)},font=\fontsize{5}{5.5}\selectfont]
\draw[black!18,line width=0.3pt] (10,-0.5) -- (10,4.5);
\draw[black!18,line width=0.3pt] (20,-0.5) -- (20,4.5);
\draw[black!18,line width=0.3pt] (30,-0.5) -- (30,4.5);
\fill[fill={rgb,255:red,148;green,103;blue,189}] (0,-0.31) rectangle (10.763,0.31);
\node[anchor=east,inner sep=1.5pt] at (0,0) {Prompt engineer};
\fill[fill={rgb,255:red,214;green,39;blue,40}] (0,0.69) rectangle (8.042,1.31);
\node[anchor=east,inner sep=1.5pt] at (0,1) {ML engineer};
\fill[fill={rgb,255:red,44;green,160;blue,44}] (0,1.69) rectangle (4.018,2.31);
\node[anchor=east,inner sep=1.5pt] at (0,2) {Data scientist};
\fill[fill={rgb,255:red,255;green,127;blue,14}] (0,2.69) rectangle (8.806,3.31);
\node[anchor=east,inner sep=1.5pt] at (0,3) {Data engineer};
\fill[fill={rgb,255:red,31;green,119;blue,180}] (0,3.69) rectangle (5.682,4.31);
\node[anchor=east,inner sep=1.5pt] at (0,4) {Data analyst};
\draw[black!70,line width=0.4pt] (0,-0.5) -- (0,4.5);
\end{tikzpicture}%

%% file: bars/hard-python-scripting.tex
\begin{tikzpicture}[x=0.08667cm,y=-5.60pt,baseline={([yshift=-1.2ex]current bounding box.north)},font=\fontsize{5}{5.5}\selectfont]
\draw[black!18,line width=0.3pt] (10,-0.5) -- (10,4.5);
\draw[black!18,line width=0.3pt] (20,-0.5) -- (20,4.5);
\draw[black!18,line width=0.3pt] (30,-0.5) -- (30,4.5);
\fill[fill={rgb,255:red,148;green,103;blue,189}] (0,-0.31) rectangle (9.217,0.31);
\node[anchor=east,inner sep=1.5pt] at (0,0) {Prompt engineer};
\fill[fill={rgb,255:red,214;green,39;blue,40}] (0,0.69) rectangle (7.142,1.31);
\node[anchor=east,inner sep=1.5pt] at (0,1) {ML engineer};
\fill[fill={rgb,255:red,44;green,160;blue,44}] (0,1.69) rectangle (8.806,2.31);
\node[anchor=east,inner sep=1.5pt] at (0,2) {Data scientist};
\fill[fill={rgb,255:red,255;green,127;blue,14}] (0,2.69) rectangle (6.593,3.31);
\node[anchor=east,inner sep=1.5pt] at (0,3) {Data engineer};
\fill[fill={rgb,255:red,31;green,119;blue,180}] (0,3.69) rectangle (6.231,4.31);
\node[anchor=east,inner sep=1.5pt] at (0,4) {Data analyst};
\draw[black!70,line width=0.4pt] (0,-0.5) -- (0,4.5);
\end{tikzpicture}%

%% file: bars/hard-performance-optimization.tex
\begin{tikzpicture}[x=0.08667cm,y=-5.60pt,baseline={([yshift=-1.2ex]current bounding box.north)},font=\fontsize{5}{5.5}\selectfont]
\draw[black!18,line width=0.3pt] (10,-0.5) -- (10,4.5);
\draw[black!18,line width=0.3pt] (20,-0.5) -- (20,4.5);
\draw[black!18,line width=0.3pt] (30,-0.5) -- (30,4.5);
\fill[fill={rgb,255:red,148;green,103;blue,189}] (0,-0.31) rectangle (7.556,0.31);
\node[anchor=east,inner sep=1.5pt] at (0,0) {Prompt engineer};
\fill[fill={rgb,255:red,214;green,39;blue,40}] (0,0.69) rectangle (12.941,1.31);
\node[anchor=east,inner sep=1.5pt] at (0,1) {ML engineer};
\fill[fill={rgb,255:red,44;green,160;blue,44}] (0,1.69) rectangle (16.746,2.31);
\node[anchor=east,inner sep=1.5pt] at (0,2) {Data scientist};
\fill[fill={rgb,255:red,255;green,127;blue,14}] (0,2.69) rectangle (2.092,3.31);
\node[anchor=east,inner sep=1.5pt] at (0,3) {Data engineer};
\fill[fill={rgb,255:red,31;green,119;blue,180}] (0,3.69) rectangle (8.137,4.31);
\node[anchor=east,inner sep=1.5pt] at (0,4) {Data analyst};
\draw[black!70,line width=0.4pt] (0,-0.5) -- (0,4.5);
\end{tikzpicture}%

%% file: bars/hard-cloud-infrastructure.tex
\begin{tikzpicture}[x=0.08667cm,y=-5.60pt,baseline={([yshift=-1.2ex]current bounding box.north)},font=\fontsize{5}{5.5}\selectfont]
\draw[black!18,line width=0.3pt] (10,-0.5) -- (10,4.5);
\draw[black!18,line width=0.3pt] (20,-0.5) -- (20,4.5);
\draw[black!18,line width=0.3pt] (30,-0.5) -- (30,4.5);
\fill[fill={rgb,255:red,148;green,103;blue,189}] (0,-0.31) rectangle (4.898,0.31);
\node[anchor=east,inner sep=1.5pt] at (0,0) {Prompt engineer};
\fill[fill={rgb,255:red,214;green,39;blue,40}] (0,0.69) rectangle (7.290,1.31);
\node[anchor=east,inner sep=1.5pt] at (0,1) {ML engineer};
\fill[fill={rgb,255:red,44;green,160;blue,44}] (0,1.69) rectangle (4.748,2.31);
\node[anchor=east,inner sep=1.5pt] at (0,2) {Data scientist};
\fill[fill={rgb,255:red,255;green,127;blue,14}] (0,2.69) rectangle (11.843,3.31);
\node[anchor=east,inner sep=1.5pt] at (0,3) {Data engineer};
\fill[fill={rgb,255:red,31;green,119;blue,180}] (0,3.69) rectangle (1.892,4.31);
\node[anchor=east,inner sep=1.5pt] at (0,4) {Data analyst};
\draw[black!70,line width=0.4pt] (0,-0.5) -- (0,4.5);
\end{tikzpicture}%

%% file: bars/hard-data-warehousing-modeling.tex
\begin{tikzpicture}[x=0.08667cm,y=-5.60pt,baseline={([yshift=-1.2ex]current bounding box.north)},font=\fontsize{5}{5.5}\selectfont]
\draw[black!18,line width=0.3pt] (10,-0.5) -- (10,4.5);
\draw[black!18,line width=0.3pt] (20,-0.5) -- (20,4.5);
\draw[black!18,line width=0.3pt] (30,-0.5) -- (30,4.5);
\fill[fill={rgb,255:red,148;green,103;blue,189}] (0,-0.31) rectangle (4.051,0.31);
\node[anchor=east,inner sep=1.5pt] at (0,0) {Prompt engineer};
\fill[fill={rgb,255:red,214;green,39;blue,40}] (0,0.69) rectangle (4.982,1.31);
\node[anchor=east,inner sep=1.5pt] at (0,1) {ML engineer};
\fill[fill={rgb,255:red,44;green,160;blue,44}] (0,1.69) rectangle (8.602,2.31);
\node[anchor=east,inner sep=1.5pt] at (0,2) {Data scientist};
\fill[fill={rgb,255:red,255;green,127;blue,14}] (0,2.69) rectangle (14.649,3.31);
\node[anchor=east,inner sep=1.5pt] at (0,3) {Data engineer};
\fill[fill={rgb,255:red,31;green,119;blue,180}] (0,3.69) rectangle (17.560,4.31);
\node[anchor=east,inner sep=1.5pt] at (0,4) {Data analyst};
\draw[black!70,line width=0.4pt] (0,-0.5) -- (0,4.5);
\end{tikzpicture}%

%% file: bars/hard-sql-database-systems.tex
\begin{tikzpicture}[x=0.08667cm,y=-5.60pt,baseline={([yshift=-1.2ex]current bounding box.north)},font=\fontsize{5}{5.5}\selectfont]
\draw[black!18,line width=0.3pt] (10,-0.5) -- (10,4.5);
\draw[black!18,line width=0.3pt] (20,-0.5) -- (20,4.5);
\draw[black!18,line width=0.3pt] (30,-0.5) -- (30,4.5);
\fill[fill={rgb,255:red,148;green,103;blue,189}] (0,-0.31) rectangle (1.946,0.31);
\node[anchor=east,inner sep=1.5pt] at (0,0) {Prompt engineer};
\fill[fill={rgb,255:red,214;green,39;blue,40}] (0,0.69) rectangle (2.938,1.31);
\node[anchor=east,inner sep=1.5pt] at (0,1) {ML engineer};
\fill[fill={rgb,255:red,44;green,160;blue,44}] (0,1.69) rectangle (6.179,2.31);
\node[anchor=east,inner sep=1.5pt] at (0,2) {Data scientist};
\fill[fill={rgb,255:red,255;green,127;blue,14}] (0,2.69) rectangle (10.646,3.31);
\node[anchor=east,inner sep=1.5pt] at (0,3) {Data engineer};
\fill[fill={rgb,255:red,31;green,119;blue,180}] (0,3.69) rectangle (10.763,4.31);
\node[anchor=east,inner sep=1.5pt] at (0,4) {Data analyst};
\draw[black!70,line width=0.4pt] (0,-0.5) -- (0,4.5);
\end{tikzpicture}%

%% file: bars/hard-data-visualization-bi.tex
\begin{tikzpicture}[x=0.08667cm,y=-5.60pt,baseline={([yshift=-1.2ex]current bounding box.north)},font=\fontsize{5}{5.5}\selectfont]
\draw[black!18,line width=0.3pt] (10,-0.5) -- (10,4.5);
\draw[black!18,line width=0.3pt] (20,-0.5) -- (20,4.5);
\draw[black!18,line width=0.3pt] (30,-0.5) -- (30,4.5);
\fill[fill={rgb,255:red,148;green,103;blue,189}] (0,-0.31) rectangle (1.680,0.31);
\node[anchor=east,inner sep=1.5pt] at (0,0) {Prompt engineer};
\fill[fill={rgb,255:red,214;green,39;blue,40}] (0,0.69) rectangle (0.895,1.31);
\node[anchor=east,inner sep=1.5pt] at (0,1) {ML engineer};
\fill[fill={rgb,255:red,44;green,160;blue,44}] (0,1.69) rectangle (5.035,2.31);
\node[anchor=east,inner sep=1.5pt] at (0,2) {Data scientist};
\fill[fill={rgb,255:red,255;green,127;blue,14}] (0,2.69) rectangle (2.209,3.31);
\node[anchor=east,inner sep=1.5pt] at (0,3) {Data engineer};
\fill[fill={rgb,255:red,31;green,119;blue,180}] (0,3.69) rectangle (15.052,4.31);
\node[anchor=east,inner sep=1.5pt] at (0,4) {Data analyst};
\draw[black!70,line width=0.4pt] (0,-0.5) -- (0,4.5);
\end{tikzpicture}%

%% file: bars/hard-distributed-computing.tex
\begin{tikzpicture}[x=0.08667cm,y=-5.60pt,baseline={([yshift=-1.2ex]current bounding box.north)},font=\fontsize{5}{5.5}\selectfont]
\draw[black!18,line width=0.3pt] (10,-0.5) -- (10,4.5);
\draw[black!18,line width=0.3pt] (20,-0.5) -- (20,4.5);
\draw[black!18,line width=0.3pt] (30,-0.5) -- (30,4.5);
\fill[fill={rgb,255:red,148;green,103;blue,189}] (0,-0.31) rectangle (0.429,0.31);
\node[anchor=east,inner sep=1.5pt] at (0,0) {Prompt engineer};
\fill[fill={rgb,255:red,214;green,39;blue,40}] (0,0.69) rectangle (2.971,1.31);
\node[anchor=east,inner sep=1.5pt] at (0,1) {ML engineer};
\fill[fill={rgb,255:red,44;green,160;blue,44}] (0,1.69) rectangle (3.470,2.31);
\node[anchor=east,inner sep=1.5pt] at (0,2) {Data scientist};
\fill[fill={rgb,255:red,255;green,127;blue,14}] (0,2.69) rectangle (9.005,3.31);
\node[anchor=east,inner sep=1.5pt] at (0,3) {Data engineer};
\fill[fill={rgb,255:red,31;green,119;blue,180}] (0,3.69) rectangle (2.029,4.31);
\node[anchor=east,inner sep=1.5pt] at (0,4) {Data analyst};
\draw[black!70,line width=0.4pt] (0,-0.5) -- (0,4.5);
\end{tikzpicture}%

%% file: bars/hard-data-pipeline-development.tex
\begin{tikzpicture}[x=0.08667cm,y=-5.60pt,baseline={([yshift=-1.2ex]current bounding box.north)},font=\fontsize{5}{5.5}\selectfont]
\draw[black!18,line width=0.3pt] (10,-0.5) -- (10,4.5);
\draw[black!18,line width=0.3pt] (20,-0.5) -- (20,4.5);
\draw[black!18,line width=0.3pt] (30,-0.5) -- (30,4.5);
\fill[fill={rgb,255:red,148;green,103;blue,189}] (0,-0.31) rectangle (0.135,0.31);
\node[anchor=east,inner sep=1.5pt] at (0,0) {Prompt engineer};
\fill[fill={rgb,255:red,214;green,39;blue,40}] (0,0.69) rectangle (2.092,1.31);
\node[anchor=east,inner sep=1.5pt] at (0,1) {ML engineer};
\fill[fill={rgb,255:red,44;green,160;blue,44}] (0,1.69) rectangle (0.928,2.31);
\node[anchor=east,inner sep=1.5pt] at (0,2) {Data scientist};
\fill[fill={rgb,255:red,255;green,127;blue,14}] (0,2.69) rectangle (3.088,3.31);
\node[anchor=east,inner sep=1.5pt] at (0,3) {Data engineer};
\fill[fill={rgb,255:red,31;green,119;blue,180}] (0,3.69) rectangle (0.546,4.31);
\node[anchor=east,inner sep=1.5pt] at (0,4) {Data analyst};
\draw[black!70,line width=0.4pt] (0,-0.5) -- (0,4.5);
\end{tikzpicture}%

%% file: TAB-HARDSKILLS-OTHER.tex
\caption{Hard skills 
	of the four reference roles (pooled).}
\label{tab:hard-skills-clusters-all}
\centering
\footnotesize
\begin{tabularx}{\textwidth}{
    p{3.2cm}
    X
    c
}
\toprule
\textbf{Cluster} & \textbf{Top Skills (count)} 
& \textbf{Freq.} \\
\midrule
\raggedright
Data Science \& Analytics Tools & 
R (2578), Power BI (2495), Spark (2488), Tableau (2479), PyTorch (1851)
 & 25.1\% \\
\raggedright
Data Warehousing \& Modeling & 
Data analysis (4279), Data modeling (2945), Data engineering (1690), Data warehousing (1517), Data governance (1012)
& 12.6\% \\
\raggedright
SQL \& Database Systems & 
SQL (11247), NoSQL (644), Postgresql (542), MySQL (416), MongoDB (371)
& 8.7\% \\
\raggedright
Performance Optimization & 
Statistical analysis (1156), Data mining (865), Predictive modeling (781), Statistical modeling (750), Computer vision (551)
& 8.2\% \\
\raggedright
Cloud Infrastructure & 
AWS (2577), Azure (1473), Kubernetes (1412), Azure Data Factory (851), Terraform (801)
& 7.6\% \\
\raggedright
Machine Learning \& AI & 
Machine learning (4268), Tensorflow (1820), Deep learning (1166), Natural language processing (760), Generative AI (551)
& 7.4\% \\
\raggedright
Python \& Scripting & 
Python (12616), Python programming (409), Matplotlib (248), Programming in Python (239), Shell scripting (180)
& 7.1\% \\
\raggedright
ETL \& Dev\-Ops & 
ETL processes (1341), DevOps (475), GitHub (354), ETL tools (339), C\# (228)
& 6.9\% \\
\raggedright
Distributed Computing & 
Databricks (1884), Pyspark (1663), Hadoop (1355), Apache Spark (862), Bigquery (698)
& 5.5\% \\
\raggedright
Data Visualization \& BI & 
Data visualization (2983), Excel (1198), Data analytics (827), Data visualization tools (701), Predictive analytics (300)
& 5.3\% \\
\raggedright
Agile \& Testing & 
A/B testing (579), Prompt engineering (285), Agile methodologies (252), Project management (130), Agile methodology (124)
& 3.6\% \\
\raggedright
Data Pipeline Development & 
Data pipelines (955), CI/CD pipelines (521), Data pipeline development (415), ETL pipelines (392), Data pipeline design (207)
& 1.9\% \\
\bottomrule
\end{tabularx}

%% file: PROMPT.tex
\noindent%
\begin{framed}
\# Step 1: Examine the job description\\
You are an expert skill extractor with specialization in analyzing technical job postings. I'm going to provide you with a job description text. Your task is to carefully analyze this text to extract structured skill information.\\[\baselineskip]

\# Context Specification\\
The input is a raw job posting that may contain various sections including job overview, responsibilities, requirements, and company information.\\[\baselineskip]

\# Step 2: Extract and categorize information\\
Follow this step-by-step process:\\
1. Read the entire job description carefully\\
2. Identify all mentioned skills, qualifications, and requirements\\
3. Categorize each identified item into the appropriate category:\\
\indent -- soft\_skills: Interpersonal and behavioral attributes (e.g., communication, teamwork)\\
\indent -- hard\_skills: Technical abilities and specific knowledge (e.g., Python, prompt engineering)\\
\indent -- qualifications: Educational background, certifications, or experience requirements (e.g., Bachelor's degree, 3+ years experience)\\
\indent -- other: Any requirements that don't clearly fit the above categories\\
4. Determine the application domain where prompt engineering will be applied. This is NOT just ``tech'' or ``AI'' but the specific industry or field where the company will apply prompt engineering (e.g., healthcare, finance, education, marketing, software development)\\[\baselineskip]

\# Output Format\\
Respond ONLY with a valid JSON object using this exact structure:\\
\{\\
\indent ``soft\_skills'': [``skill1'', ``skill2'', ...],\\
\indent ``hard\_skills'': [``skill1'', ``skill2'', ...],\\
\indent ``qualifications'': [``qual1'', ``qual2'', ...],\\
\indent ``job\_domain'': ``The specific industry or application area where prompt engineering will be applied'',\\
\indent ``other'': [``item1'', ``item2'', ...]\\
\}\\[\baselineskip]

Important guidelines:\\
-- Extract skills as concise noun phrases (1--4 words typically)\\
-- Separate distinct skills (don't combine multiple skills into one entry)\\
-- Format consistently using lowercase\\
-- Include ALL relevant skills mentioned, beyond prompting\\
-- For job\_domain, identify the SPECIFIC industry or application area where the company operates and will apply prompt engineering (NOT just ``technology'' or ``AI'')\\
-- If multiple domains are mentioned, list the primary domain or focus\\
-- The output must be valid JSON (double quotes, commas between items)\\
-- Contain only the JSON object, no additional text
\end{framed}